\documentclass[11pt, a4paper]{article}
\usepackage{jheparxiv}
\usepackage[utf8]{inputenc}
\usepackage{amsmath}
\usepackage{comment}
\newcommand{\be}{\begin{equation}}
\newcommand{\ee}{\end{equation}}
\usepackage{amsmath,bm}
\usepackage[all]{xy}
\usepackage{tikz-cd}
\usepackage{dsfont}
\usepackage{bm}
\usepackage{amsfonts}
\usepackage{appendix}
\usepackage{amssymb}
\usepackage{latexsym}
\usepackage{float}
\usepackage{mathrsfs}
\usepackage{braket}	
\usepackage{graphicx}
\usepackage{color}
\usepackage{xcolor}
\usepackage{slashed}
\usepackage{mathtools}
\usepackage{sujai}
\usepackage{breqn}
\usepackage[all]{xy}
\usepackage{tikz-feynman}

\begin{document}
\newcommand{\RNum}[1]{\uppercase\expandafter{\romannumeral #1\relax}}
\tikzfeynmanset{compat=1.1.0}
\subheader{\hfill \texttt{}}
\newcommand{\com}[1]{ \textcolor{red}{(#1)}}
\newcommand{\blue}[1]{ \textcolor{blue}{#1}}

\title{Double Copy in AdS$_3$ from Minitwistor Space }
\author[a]{Cameron Beetar,}
\author[b]{Mariana Carrillo González,}
\author[b,c]{Sumer Jaitly}
\author[b]{and Théo Keseman}
\affiliation[a]{The Laboratory for Quantum Gravity \& Strings, Department of Mathematics and Applied Mathematics, University of Cape Town, Cape Town, South Africa}
\affiliation[b]{Theoretical Physics Group, Blackett Laboratory \\
Imperial College London, SW7 2AZ, United Kingdom}
\affiliation[c]{Scuola Normale Superiore, Piazza dei Cavalieri 7, 56126, Pisa, Italy}
\emailAdd{btrcam001@myuct.ac.za}
\emailAdd{m.carrillo-gonzalez@imperial.ac.uk}
\emailAdd{sumer.jaitly@sns.it}
\emailAdd{theo.keseman17@imperial.ac.uk}
\abstract{The double copy relates gravitational theories to the square of gauge theories. While it is well understood in flat backgrounds, its precise realisation around curved spacetimes remains an open question. In this paper, we construct a classical double copy for cohomology class representatives in the minitwistor space of hyperbolic spacetimes. We find that the realisation of a physical double copy requires that the masses of the different spinning fields are not equal, contrary to the flat space prescription. This leads to a position-space double copy for bulk-to-boundary propagators. We also show that in coordinate space, this implies the Cotton double copy for waves and warped black holes of Topologically Massive Gravity. We show that these are exact double copy relations by constructing their Kerr-Schild metrics and also analysing the Kerr-Schild double copy. Furthermore, we find that near the boundary the double copy relates the dual CFT currents. }

\maketitle

%%%%%%%%%%%%%%%%%%%%%%%%%%
%%%%%%%%%%%%%%%%%%%%%%%%%%%%%

\section{Introduction}
While gravitational and gauge theories may seem to be completely unrelated theories, it is possible to relate the former to the {\it square} of the latter. This is done via the double copy, which was first realised as a relation between closed and open string amplitudes \cite{Kawai:1985xq} and in the low energy limit gives the schematic relation: gravity $\sim$ (gauge theory)$^2$. It is now understood that these theories satisfy the so-called colour-kinematics duality \cite{Bern:2008qj, Bern:2010ue} in which the kinematic and colour factors appearing in gauge theory scattering amplitudes, both obey the same algebraic identity. For the colour factors this identity simply follows from the Jacobi identity whereas the origin of the kinematic identity is only well understood in a limited number of cases \cite{Monteiro:2011pc, Monteiro:2013rya,Cheung:2016prv,Chen:2019ywi,Borsten:2023ned,Ben-Shahar:2021zww,Ben-Shahar:2022ixa,Brandhuber:2021bsf,Brandhuber:2022enp,Fu:2016plh,Reiterer:2019dys,Tolotti:2013caa,Ferrero:2020vww,Borsten:2019prq,Borsten:2020zgj,Borsten:2020xbt,Borsten:2021hua,Diaz-Jaramillo:2021wtl,Bonezzi:2022bse,Bonezzi:2022yuh,Bonezzi:2023lkx,Bonezzi:2024dlv,Armstrong-Williams:2024icu, Borsten:2022vtg, Borsten:2023paw, Borsten:2023reb}.

The BCJ double copy has been proven at tree level and while many examples have been computed to high loop orders, finding colour-dual kinematic factors is a difficult task \cite{Adamo:2022dcm}. Nevertheless, generalised double copy methods \cite{Bern:2017yxu,Bern:2017ucb} and a novel loop-integrand basis \cite{Bern:2024vqs} allow one to leverage the double copy structure to compute loop integrands without the knowledge of these colour-dual numerators. Due to the relative simplicity of calculations in gauge theories as compared to those in gravitational theories, the double copy permits otherwise almost intractable gravitational loop computations which shed light on ultraviolet divergences \cite{Bern:2023zkg}. 

It is now well known that double copy relations are not exclusive to Yang-Mills theory and Einstein gravity \cite{Adamo:2022dcm}. Higher derivative corrections can also be included, as well as scalar effective field theories and some theories with massive mediators; for a recent compilation of double copies see \cite{Bern:2022wqg}. Regarding massive gauge and gravitational theories, it has been shown that while at 4-points one can construct dRGT massive gravity through the double copy, the construction fails at 5-points \cite{Momeni:2020vvr,Johnson:2020pny}. On the other hand, theories arising from a Kaluza-Klein reduction will have a physical double copy at all points \cite{Momeni:2020hmc,Johnson:2020pny,Gonzalez:2022mpa}. An interesting example arises in three dimensions (3d) and corresponds to topologically massive theories \cite{Gonzalez:2021bes,Gonzalez:2021ztm,Moynihan:2020ejh,Moynihan:2021rwh,Burger:2021wss,Hang:2021fmp,Hang:2021oso}. These theories propagate one massive helicity mode and are gauge invariant. It has been shown that they satisfy a physical double copy at 5-points given the existence of a generalised BCJ relation for their scattering amplitudes \cite{Gonzalez:2021bes}. In this paper, we will explore further double copy relationships arising in these theories.

The double copy structure is not limited to scattering amplitudes, but can extend to relationships between classical solutions of gauge theory and gravity. These include perturbative and exact solutions \cite{White:2024pve}. For example, the Kerr-Schild double copy \cite{Monteiro:2014cda} relates exact gravitational solutions with a Kerr-Schild metric to their single copy, which corresponds to a Maxwell field. Three-dimensional constructions can be found in \cite{CarrilloGonzalez:2019gof,Alkac:2022tvc,Alkac:2021seh,Gumus:2020hbb}. Similarly, the Weyl double copy \cite{Luna:2018dpt} relates the Weyl tensor to the square of the field strength of the gauge theory. In 3d, the analogue of the Weyl double copy is the Cotton double copy which has been analysed in flat spacetimes in \cite{CarrilloGonzalez:2022mxx,Emond:2022uaf,CarrilloGonzalez:2022ggn}. The Weyl and Cotton double copy have furthermore been discovered to arise from a particularly simple multiplicative structure on \textit{twistor space} \cite{White:2020sfn}. Twistor space is a projective space that encodes the conformal structure  of  spacetimes in terms of holomorphic functions in a non-local way \cite{Adamo:2017qyl}. Of particular relevance for the classical double copy is the existence of an isomorphism between cohomology classes and solutions to the free equations of motion of irreducible representations in the corresponding spacetime. While in 4d this is restricted to the zero-rest-mass equations, in 3d we can have a non-zero mass. The existence of classical double copy relations in flat space from twistor space has been explored in \cite{Farnsworth:2021wvs,CarrilloGonzalez:2022ggn,Luna:2022dxo,Chacon:2021lox,Chacon:2021wbr,Armstrong-Williams:2023ssz,White:2020sfn}. 

It has also been shown that some of these classical double copy relationships hold on certain curved backgrounds \cite{carrillogonzalez2018classical,Farnsworth:2023mff,Liang:2023zxo,Alkac:2023glx,Han:2022mze,Prabhu:2020avf,Bahjat-Abbas:2017htu,Ilderton:2024oly,Garcia-Compean:2024uie,Chacon:2024qsq,Chawla:2022ogv}, but whether the double copy can be realised in generic situations around curved spacetimes is an open question.  In this paper, we explore this question in three-dimensional hyperbolic spacetimes from twistor space. The Penrose transform in this case gives an isomorphism between cohomology classes and solutions to the equations of motion of free massive irreducible representations in hyperbolic space \cite{tsai}. Furthermore, since the curvature tensors of topologically massive theories satisfy these equations of motion without sources, the Penrose transform allows us to construct classical solutions of these theories. In general, these will correspond to linearised solutions, except in special cases where the spacetimes have a Kerr-Schild metric. In these special cases, the linearised solutions are exact. 

We will show that for Petrov type N spacetimes, we can construct a double copy for wave solutions, but also a double copy for bulk-to-boundary propagators. It has previously been shown that these bulk-to-boundary propagators can be obtained from twistor space in AdS$_5$ in \cite{Adamo:2016rtr}, and for scalars in AdS$_3$ in \cite{Bu:2023cef}. Besides being of interest in the construction of correlation functions in AdS$_3$, this is also relevant in the context of flat space holography. AdS$_3$ correlation functions are useful in the context of celestial holography since one can foliate Minkowski with EAdS slices \cite{Atanasov:2021oyu,Casali:2022fro,Iacobacci:2022yjo,deBoer:2003vf,Sleight:2023ojm,Melton:2023bjw,Melton:2024jyq,Melton:2024gyu}. Hence, minitwistor space has been used in celestial holography approaches \cite{Bu:2023cef,Bu:2023vjt,Bu:2024cql}. Related twistor constructions have been useful in ambitwistor strings on an AdS$_3$ target space \cite{Bhat:2021dez,Roehrig:2020kck,McStay:2023thk}. 

When it comes to Petrov type D spacetimes, which include black hole solutions, the flat space explorations of \cite{CarrilloGonzalez:2022mxx} showed that the double copy does not hold in position space. The shortcoming of this analysis is that it is only linear, since there are no exact type D solutions of Topologically Massive Gravity (TMG) that are asymptotically flat. Hence, in this paper, we will explore exact type D solutions which correspond to warped AdS spacetimes \cite{Chow:2009km}. Furthermore, these warped spacetimes locally represent 3d black hole solutions \cite{Anninos:2008fx} which are the warped analogue of BTZ black holes \cite{Banados:1992wn}. 
Warped AdS spacetimes have also been largely explored in the context of holography \cite{Chen:2013aza,Chen:2009hg,Detournay:2012pc,Anninos:2013nja,Castro:2015csg,Jensen:2017tnb,Song:2017czq,Ghodrati:2019bzz,Aggarwal:2022xfd,Aggarwal:2019iay} and correspond to the near-horizon limit of extremal Kerr black holes at a fixed polar angle \cite{Bardeen:1999px,Bengtsson:2005zj}.

This paper is structured as follows: in Sections \ref{sec:TMTheories} and \ref{sec:HyperbolicST}, we give the necessary background on topologically massive theories and the twistor correspondence between hyperbolic spacetime and minitwistor space. In Section \ref{sec:DC}, we present our mass prescription for the twistor double copy, which we apply to type N fields in Section \ref{sec:typeN} and type D fields in Section \ref{sectiontyped}. In the last two sections, we briefly turn to the (linearised) twistor double copy of more general algebraic types and to the case of massless fields. 

In Appendix \ref{ap:conventions} we state our conventions regarding index notation for tensor and spinor objects and their contractions. Appendix \ref{ap:scalar} details how the Penrose transform solves the scalar equation of motion on hyperbolic space. Appendix \ref{app:homogeneous} reviews how hyperbolic space and its associated minitwistor space can be viewed as homogeneous spaces, and subsequently how homogeneous functions of specific scaling weight arise. Appendix \ref{ap:rep} summarises how Čech and Dolbeault representatives are related. In Appendix \ref{app:dimred}, we explore cases in which minitwistor (3d) representatives can be obtained from four-dimensional twistor space representatives via the Mellin transform -- and where this can fail. Finally, Appendix \ref{app:petrov} recaps the Petrov classification of the Cotton spinor.

%%%%%%%%%%%%%%%%%%%%
\section{Topologically massive theories}
\label{sec:TMTheories}
In what follows, we will use spinor notation in which a tensor index $\mu$ (or $A$ where vielbeins are used) is replaced by a pair of spinor indices $\alpha, \beta$ via a two-to-one homomorphism. In four dimensions (4d), we work in Lorentzian signature $(+,-,-,-)$ and use the homomorphism between the proper Lorentz group $SO(1,3)$ and $SL(2,\mathbb{C})$. The 4d embedding space coordinates will have dimensions of length. Meanwhile, in 3d we will fix the background to real hyperbolic space, which we also refer to as Euclidean Anti-de Sitter (EAdS), with signature $(-,-,-)$. The 3d hyperbolic space coordinates will be dimensionless. Locally, we can work in an orthonormal frame defined via a basis of vielbeins, and we can use the homomorphism between $SO(3)$ and $SU(2)$ (or $SO(1,2)$ and $SL(2,\mathbb{R})$ in Lorentzian signature) to go to the spinor basis. This is described in detail in Appendix \ref{ap:conventions}. With this choice of conventions\footnote{We take the Euclidean signature of the conventions in \cite{Clement:1994sb}.}, we have that the Ricci scalar is given by $R= -6\Lambda=\frac{6}{L^2}$. Similarly, the dimensionless masses are defined as  $\widetilde m \equiv m L$, where $L$ is the EAdS length scale. 

\subsection{Topologically massive Yang-Mills} \label{sec:TMYM}
We start by reviewing topologically massive Yang-Mills theory (TMYM). In a three-dimensional curved spacetime its action is \cite{Deser:1981wh}\footnote{The $i$ factor in front of the Chern-Simons term is due to the fact that it is time-reversal odd and we are working in Euclidean signature.}
\begin{equation}\label{tmym action}
    S_{\tmym}= \frac{1}{\tilde{g}^2}\int \rd^3x\sqrt{-g}\,\text{tr}\left(-\frac{1}{2} F^{\mu \nu}F_{\mu\nu} +2 \tilde{g}  A^{\mu }J_{\mu } - \frac{i\mu_V}{2} \varepsilon^{\mu \nu \rho} \left(F_{\mu\nu} -\frac{2}{3} A_{\mu}A_{\nu}\right)A_{\rho}\right) \ ,
\end{equation}
where the field strength of the dimensionless non-Abelian gauge field, $A_{\mu}=A_{\mu}^aT^a$, is $F_{\mu\nu}= \partial_{\mu}A_{\nu}-\partial_{\nu}A_{\mu}-i [A_{\mu},A_{\nu}]$ and the generators of the gauge group satisfy $[T^a,T^b]=i \widetilde{f}^{abc}T^c$, with $\widetilde{g}^2$ the coupling strength with units of mass, $\mu_V $ the mass of the gauge field, and the curved space epsilon tensor given by $\varepsilon_{\mu \nu \rho}=\sqrt{-g}\epsilon_{\mu \nu \rho}$. Under gauge transformations, the Chern-Simons term shifts by a total derivative plus a Wess-Zumino-Witten term. Selecting suitable boundary conditions, the former vanishes and the latter can give a non-zero contribution when considering large gauge transformations. If the Wess-Zumino-Witten term gives a nonzero winding number, the path integral will still be gauge invariant as long as the Chern-Simons level $k$, related to the mass as $k=2 \pi \mu_V/ \tilde{g}^2$, is an integer\footnote{One can take the point of view that this theory should be invariant under large gauge transformations in all manifolds, which immediately leads to the quantisation of the level when considering the three-sphere manifold and working with gauge groups given by compact connected simple Lie groups whose third homotopy group is non-trivial \cite{Dunne:1998qy}. The contribution of the Wess-Zumino-Witten term when considering hyperbolic manifolds is more complicated and can be highly dependent on boundary conditions and regularisation. Similar arguments apply to the gravitational Chern-Simons term.}. 

This theory describes a single, parity-violating, massive spin\footnote{In 3 dimensions, spin is a rotational pseudoscalar.} one local degree of freedom as shown in \cite{Deser:1981wh}. Changing the sign in front of the Chern-Simons term, or equivalently changing the sign of $ \mu_V$, simply changes the helicity of the massive gluons. In Eq.~\eqref{tmym action} positive mass implies positive helicity. The Bianchi identity and the equations of motion are then
\begin{equation}\label{tmym}
\begin{aligned} 
    D_{[\mu}F_{\nu \rho]}&\,=0 \ , \\
    D_{\mu}F^{\mu \nu}- i \frac{\mu_V}{2} \varepsilon^{\nu\rho\gamma}F_{\rho \gamma}&\,=  \tilde{g}J^{\mu} \ .
\end{aligned}
\end{equation}
When considering the double copy of classical solutions, we will work with linearised solutions, which can be exact in some specific cases. For this purpose, we consider the ansatz $A^{\mu a}=c^a A^{\mu}$ and $J_{\mu a}= c_a J_{\mu}$, such that the equations of motion linearise and reduce to the topologically massive Maxwell's equations. Furthermore, when obtaining solutions from the twistor space, we will set $J_{\mu a}=0$. Renaming the mass $ m_V=i \mu_V$ for later convenience, the equations can be restated in spinor notation as \cite{CarrilloGonzalez:2022mxx}
\begin{equation}\label{tmymspinor}
\nabla_{\alpha}^{\phantom{a}\gamma}f_{\beta \gamma}=- m_V f_{\alpha\beta} \ ,
\end{equation}
where $f_{\alpha\beta}=\sigma^A_{\alpha \beta} e^{\mu}_A f_{\mu}$ is the spinor of the dual field strength, $f_{\mu}=\frac{1}{2} \varepsilon_{\mu\nu\rho}F^{\nu \rho}$, $e^{\mu}_A$ are the vielbeins (which we take to be dimensionless so that $\epsilon_{\alpha \beta}$ is dimensionless and $[\nabla_{\alpha}^{\phantom{a}\gamma}]=L^{-1}$), and $\sigma^A_{\alpha \beta}$ can be found in the Appendix~\ref{ap:conventions}. Note that $\nabla_{[\alpha}^{\phantom{a}\gamma}f_{\beta]\gamma}=0$ arises as an integrability condition, that is, it is required by consistency of the equations of motion.

\subsection{Topologically massive gravity}
We now give a short review of topologically massive gravity~\cite{Deser:1981wh}. Including a cosmological constant term, its action is given by
\begin{equation}\label{tmg}
    S_{\tmg} = \frac{1}{\kappa^2}\int \rd^3x \sqrt{-g}\left ( -R -2 \Lambda -\frac{i}{2 \mu_G} \varepsilon^{\mu\nu\rho}\Gamma^{\alpha}_{\mu\sigma}\left(\partial_{\nu}\Gamma^{\sigma}_{\alpha\rho} +\frac{2}{3}\Gamma^{\sigma}_{\nu\beta}\Gamma^{\beta}_{\rho\alpha}\right)\right) \ ,
\end{equation}
where $\kappa^2 = 16 \pi G$, $\Lambda=-\frac{1}{L^2}$, %$\Lambda=-1$
and $\mu_G$ is the mass
%$\widetilde \mu_G$ is the dimensionless mass 
of the graviton. Contrary to Einstein's theory in three dimensions, this theory is non-trivial in the bulk and possesses a single, parity-violating, massive spin 2 bulk degree of freedom. As in the gauge theory case, the sign of the mass is related to the helicity of the massive graviton. The choice of the sign of the Einstein-Hilbert term can lead to different negative energy states as reviewed in \cite{Alkac:2017vgg}. One possibility is to take the sign of the Ricci scalar to be the usual sign in Einstein's General Relativity, in which case the BTZ and warped AdS black holes have positive energy, but the massive graviton is a ghost~\cite{Deser:1981wh,Carlip:2008jk,Li:2008yz}. In the context of AdS/CFT, both of these choices lead to a non-unitary CFT on the boundary \cite{Skenderis:2009nt} \footnote{A possibility to have a well-defined boundary theory is to consider the critical point $\mu_G L=1$ \cite{Li:2008dq,Giribet:2008bw,Maloney:2009ck}.}. The second possibility is to take the opposite sign of $R$ so that gravitons propagate physically in the bulk but black holes have generically negative energy (and should therefore be excluded from the space of physical solutions). We picked the first option here since we are interested in black hole solutions and we are not currently looking at the scattering of bulk gravitons. \\

The equations of motion of TMG are given by
\begin{equation}\label{eomtmg}
    G_{\mu\nu} - \Lambda g_{\mu\nu}+\frac{i}{ \mu_G}C_{\mu\nu}=0 \ ,
\end{equation}
where $G_{\mu\nu}$ is the Einstein tensor and $C_{\mu\nu}$ the Cotton (also referred to as Cotton-York) tensor.  The Weyl tensor vanishes identically in three dimensions, and the conformal structure is now given by the Cotton tensor defined as
\begin{equation}
    C^{\mu\nu}= \varepsilon^{\mu\alpha\beta}\nabla_{\alpha}\left(R^{\nu}_{\beta}-\frac{1}{4}g^{\nu}_{\beta}R\right) = \varepsilon^{\mu\alpha\beta}\nabla_{\alpha}S^{\nu}_{\beta} \ , \label{eq:cottonDef}
\end{equation}
where $S^{\nu}_{\beta}$ is traceless Ricci tensor and the second equality only holds when the Ricci scalar is constant. By the Bianchi identity, the Cotton tensor is symmetric and traceless. Rewriting $ m_G= i  \mu_G$, the Bianchi identity and equations of motion can be written in spinor notation as \cite{CarrilloGonzalez:2022mxx}
\begin{equation}\label{tmgspinor}
    \nabla_{\alpha}^{\phantom{a}\epsilon}C_{\beta \gamma \delta \epsilon}
        = -  m_{G}C_{\alpha \beta \gamma \delta} \ .
\end{equation}
As in the gauge theory case, $\nabla_{[\alpha}^{\phantom{aa}\epsilon}C_{\beta \gamma \delta] \epsilon} =0$, arises as an integrability condition. Note that we will work on a fixed AdS background, which means that all our results will be linearisations to the full solutions of topologically massive gravity. In some specific scenarios, this will correspond to exact solutions, not just linearisations of the theory.

\subsection{Topologically massive higher spin theories and irreducible representations of EAdS} \label{sec:higherspin}
Similar to Einstein's gravity, massless higher-spin theories have no propagating degrees of freedom in three dimensions \cite{Campoleoni:2024ced,10.1063/1.525524,Kessel:2017mxa}. However, massive, parity-violating higher-spin theories can have a propagating degree of freedom in the bulk \cite{Vasiliev:1992gr,Barabanshchikov:1996mc,Bagchi:2011vr,Kuzenko:2018lru,Chen:2011vp,Boulanger:2000rq,Kessel:2018zqm,Chen:2011yx}. Their (linearised) equations of motion in AdS in spinor notation are
\begin{equation}\label{higherspineom}
\begin{aligned}
    \nabla^{\beta \gamma} \Psi_{\beta \gamma \alpha_1\ldots\alpha_{2s-2}}&\,=0 \ , \\
    \nabla^{\beta}_{\phantom{a}(\alpha_1}\Psi_{\alpha_2\ldots\alpha_{2s})\beta}&\,= -m_{(s)} \Psi_{\alpha_1\ldots\alpha_{2s}} \ ,
\end{aligned}
\end{equation}
where $\Psi_{\alpha_1\ldots\alpha_{2s}}= \Psi_{(\alpha_1\ldots\alpha_{2s})}$, and where after Wick rotating as previously we define the mass as $m_{(s)}=i \mu_{(s)}$.
From now on, we will always refer to the mass as $m$ rather than $\mu$ irrespective of the signature.
These correspond to the equations of motion of massive fields in the irreducible representations of the symmetry algebra of EAdS. More precisely, we can work in the $\mathfrak{so}(1,3)$ basis in which the representations, $D(\Delta,s)$, are labelled by the conformal dimension $\Delta$ related to the dimensionless mass of the field
\begin{equation}
    \widetilde{m}=m L \ ,
\end{equation}
as
\begin{equation}\label{deltam}
    \Delta_\pm= 1 \pm \widetilde{m}_{(s)} \ , 
\end{equation}
and the spin $s$. This relationship can be seen by defining the operator
\begin{align}
(\mathcal{D}_\pm \Psi)_{\alpha_1 \ldots \alpha_{2 s}} \equiv \left(\nabla_{\alpha_1}^\beta \pm m_{(s)} \epsilon _{\alpha_1}^\beta \right)\Psi_{\beta \ldots \alpha_{2 s}} \ ,
\end{align}
and applying it twice on the higher-spin field, thus obtaining 
\begin{equation} 
2(\mathcal{D}_{-} \mathcal{D}_{+} \Psi)_{\alpha_1 \ldots \alpha_{2 s}}=\nabla^{\alpha\beta}\nabla_{\alpha\beta}\Psi_{\alpha_1\ldots\alpha_{2s}}+\frac{2}{L^2}\left( s- \widetilde{m}_{(s)}^2+1\right)\Psi_{\alpha_1\ldots\alpha_{2s}}=0 \ .
\label{boxgeneralspinor}
\end{equation}
On the other hand, one can rewrite the LHS in terms of EAdS Killing vectors, which together with the expressions for the Casimirs of the Euclidean conformal group lead to
\begin{equation} \label{eq:boxPsi}
(\nabla^{\alpha\beta}\nabla_{\alpha\beta} - 2M_{(s)}^2)\Psi_{\alpha_1\ldots\alpha_{2s}}=0 \ .
\end{equation}
where in our conventions $\Box=\nabla^{\alpha\beta}\nabla_{\alpha\beta}/2$ and 
\begin{equation}
    M_{(s)}^2=\frac{\Delta(\Delta-2)-s}{L^2} \ .
\end{equation}
When $s=0$, Eq.~\eqref{higherspineom} does not exist, but Eq.~\eqref{eq:boxPsi} still holds. Combining the above expressions one obtains Eq.~\eqref{deltam}. Alternatively, it can be convenient to work in the $\mathfrak{sl}(2,\mathbb{C})$ basis where now the representations are labeled by their holomorphic and anti-holomorphic conformal weights $h,\bar{h}$. The irreducible representations in the two basis are related as \cite{Kessel:2018zqm}
\begin{equation}\label{sl2basis}
    D(\Delta,s)=(\mathcal{D}_+(h), \mathcal{D}_-(\bar{h})) \ ,
\end{equation}
where
\begin{equation}
    \begin{aligned}
        \Delta&\,=h+\bar{h} \ ,\\
        s &\,=\vert h-\bar{h} \vert\ ,
    \end{aligned}
\end{equation}
and the helicity is given by
\begin{equation}
    \eta=h-\bar{h} \ .
\end{equation}
The equation for the conformal dimension in terms of the mass in Eq.~\eqref{deltam} can now be explicitly inverted and for $s\neq0$ gives
\begin{equation}
\widetilde{m}_{(s)}=\text{Sign}(\eta)(\Delta - 1) \ ,
\label{eq:MassHelicityDelta}
\end{equation}
which makes explicit the relation mentioned earlier between the sign of the mass and the helicity of the particle.
The representations on the right-hand side (RHS) of Eq.~\eqref{sl2basis} are unitary if and only if $h \geq 0, \Bar{h} \geq 0$ which is consistent with the unitarity requirement (in Lorentzian signature\footnote{In Euclidean signature, this corresponds to reflection positivity. \cite{Penedones:2016voo}}) on the left-hand side (LHS) of Eq.~\eqref{sl2basis} that 
\begin{align}\label{unitaritybound}
        \Delta &\,\geq s \ , \quad s \neq 0 \ ,\\
        \Delta &\,\geq 0 \ , \quad s=0 \ .
\end{align}
Topologically massive theories in (E)AdS, including their higher-spin versions,  describe particles in the irreducible representation $(\mathcal{D}_+(h), \mathcal{D}_-(\bar{h}))$, which is explicitly parity odd for $s \neq 0$, contrarily to the Fierz-Pauli equations in AdS$_3$ which are described by $(\mathcal{D}_+(h), \mathcal{D}_-(\bar{h})) \oplus (\mathcal{D}_+(\bar{h}), \mathcal{D}_-(h))$ and are therefore parity invariant.

%%%%%%%%%%%%%%%%%%%%
\section{Hyperbolic spacetimes and their (Mini)twistor spaces}
\label{sec:HyperbolicST}
We are interested in double copy relations arising in theories defined on three-dimensional hyperbolic background spacetimes. These spacetimes can all be described via embeddings in four-dimensional complexified Minkowski spacetime, $\mathbb{M}_\C$ (also referred to as `embedding space'). Complexifying the spacetime removes the notion of signature and allows one to continue between different signatures by imposing so-called `reality conditions' that correspond to specific coordinates taking either purely real or purely imaginary values.

%%%%%%%%%%%%%%%%%%%%
\subsection{Complex hyperbolic space}
The three-dimensional complex hyperbolic space $\h$ can be defined by starting from four-dimensional complexified Minkowski spacetime with coordinates $X^\mu=(T,X,Y,Z)$ and line element
\begin{equation}
    \rd s^{2}=\eta_{\mu\nu}\rd X^\mu\rd X^\nu=\rd X\cdot \rd X= \rd T^2-\rd X^{2}-\rd Y^{2}-\rd Z^{2}\,,
\end{equation}
by removing the quadric $X \cdot X= 0$, and projectivising the coordinates so that $X^\mu\sim r X^\mu$ for any non-zero complex number $r\in\C^*$. The projectivised coordinates, $[X^{\mu}]$ are distinguished from the standard Minkowski coordinates by square brackets. In other words
\begin{equation}\label{HC}
    \h=\cp^3\setminus\{X\cd X=0\}\,.
\end{equation}
To get a non-projective coordinate system for hyperbolic space one must pick a single representative from each equivalence class, for instance, the representative $ x^{\mu}$ satisfying
 $ x\cdot x=1$, which can always be achieved by a scaling of $X^\mu$, as $X\cdot X\neq0$ by definition. This leaves a residual $\mathbb{Z}_2$ equivalence between $ x$ and $- x$ implying
\begin{equation}\label{hyperbolic}
    \h \cong \left\{ x^\mu \in \mathbb{C}^4\,|\, x\cdot x=1\right\}/\,\mathbb{Z}_2\,.
\end{equation}
This returns us to the more familiar formulation of complex hyperbolic space as an embedding in complexified Minkowski space. From this point onwards, we choose $x^\mu$ to be the representative for $[X^\mu]$ given by 
\begin{equation} \label{eq:CoordHyperbolic}
    x^\mu=\frac{X^\mu}{\sqrt{X\cdot X}} \ .
\end{equation}
The metric induced on the embedded hyperbolic space can be expressed in terms of the embedding space coordinates as
\begin{equation}\label{embedded}
    \rd s^2 =\eta_{\mu\nu}\, \rd \left(\frac{X^\mu}{\sqrt{X\cdot X}}\right)\rd\left(\frac{X^\nu}{\sqrt{X\cdot X}}\right).
\end{equation}

As the coordinates are complex-valued, there is no longer any notion of metric signature, nor any notion of dimensionful length. To relate $\h$ to real hyperbolic spacetimes with the usual dimensionful coordinates and well-defined signatures we must both introduce a real length scale $L\in\mathbb{R}^+$ by making the replacement $x^\mu\xr x^\mu/L$ as well as choosing a reality condition. For example, if we restrict all the $x^\mu$ coordinates to be real-valued, we get the three-dimensional surface $ t^2- x^2- y^2- z^2=L^2$ embedded in four-dimensional Minkowski spacetime, i.e. an Euclidean Anti-de Sitter spacetime\footnote{The embedded spacetime is \textit{Euclidean} Anti-de Sitter as the embedding space has metric signature $(+,-,-,-)$. One can obtain Lorentzian Anti-de Sitter by considering the embedding space with split signature $(+,+,-,-)$.} with length scale $L$. 

%%%%%%%%%%%%%%%%%%%%
\subsection{(Mini)twistor theory of complex hyperbolic space}
We shall focus on the \textit{minitwistor space} of the complex hyperbolic space described above. This space is an example of an Einstein-Weyl space and the minitwistor space $\MT$ corresponding to an Einstein-Weyl space $W$ is defined as the space of totally geodesic null hypersurfaces in $W$ \cite{Hitchin:1982vry,Jones:1985pla}. The minitwistor space of $\h$, denoted $\MT$ is \cite{tsai, Seet:2024vmh}
\begin{equation} \label{eq:MT}
    \MT = \cp^1\times \cp^1\,.
\end{equation}
It is defined through the double fibration
\begin{equation*}
\xymatrix{
 & \mathbb{F} \ar[ld] \ar[rd] & \\
 \mathbb{MT} & & \mathbb{H}} 
\end{equation*}
where $\mathbb{F} = \{ (p,x)\vert p\in \mathbb{MT}, \  x \in \mathbb{H},  \ x \in T_{p}\mathbb{MT} \}$, is the correspondence space. On the minitwistor space we can use two pairs of projective coordinates, $([\mu^{\dot\alpha}],[\lambda_{\alpha}])\in\MT$ which (in contrast to the twistor space of four-dimensional Minkowski spacetime) have independent projective equivalence relations, i.e. $([\mu^{\dot\alpha}],[\lambda_{\alpha}])\sim([u\mu^{\dot\alpha}],[v\lambda_{\alpha}])$, for independent $u,v\in\C^*$. Functions on minitwistor space can thus have independent homogeneities (scaling weight) with respect to either pair of coordinates. We say the function $f(\mu^{\dot\alpha},\lambda_{\alpha})$ is homogeneous of weight $(p,q)$ if
\begin{equation}\label{scaling}
    f(u\mu^{\dot\alpha},v\lambda_\alpha) = u^p v^q f(\mu^{\dot\alpha},\lambda_\alpha)\,.
\end{equation}
For details on how these scalings arise see Appendix~\ref{app:homogeneous}. Note that one can also define $\MT$ as the points in $\cp^3$ such that $X\cdot X=0$ since we can parameterise these points as $X^{\alpha\dot{\alpha}}=\lambda^a\mu^{\dot\alpha}$. Thus, $\MT$ corresponds to the conformal boundary of $\mathbb{H}$.

As usual for a twistor correspondence, the relationship between the hyperbolic space and minitwistor space is non-local in nature. More precisely, a point in spacetime corresponds to a linearly holomorphically embedded Riemann sphere ($\cp^1$) in the minitwistor space. This embedding is described by the \textit{incidence relation}
\begin{equation}
    \mu^{\dot\alpha} = X^{\alpha\dot\alpha}\lambda_{\alpha} \ ,
    \label{eq:incidence}
\end{equation}
such that a point with coordinate $x^{\alpha\dot\alpha}$ (the representative of class $[X^{\ada}]$) is mapped to the subset of minitwistor space given by $([x^{\alpha\dot\alpha}\lambda_\alpha],[\lambda_\alpha])$. The notation $\mu_{x}^{\dot\alpha}$ will be used for expressions where the incidence relations are obeyed by $\mu^{\dot\alpha}$. Similarly, a point $p$ in minitwistor space maps to points in complex hyperbolic space which lie on $T_{p}\mathbb{MT}$ so it corresponds to a geodesic null plane.\\

The main result we shall be using in this work is a version of the \textit{holomorphic Penrose transform} that applies to complex Einstein-Weyl spaces (including complex hyperbolic space) and their corresponding minitwistor spaces. Relating cohomological data on the minitwistor space to solutions of particular equations of motion on the complex spacetime, the Penrose transform is the following isomorphism \cite{alma990162587480107026,tsai}, for $s\geq\frac12$
\begin{equation}\label{nonzerospin}
    H^{1}(U,\cO(-s-1-\widetilde{m}\,|\,-s-1+\widetilde{m}))\cong\left\{\nabla_{\h\,\alpha_1}^{\phantom{\h\alpha}\beta}\Psi_{\beta\alpha_2\ldots\alpha_{2s}}=-\widetilde{m}\Psi_{\alpha_1\ldots\alpha_{2s}}\right\} \ ,
\end{equation}
where $U$ is an open subset of $\MT$ \footnote{This is required so that the fibers in the double fibration of the correspondence space under Eq.~\eqref{eq:MT}
are contractible and the isomorphism defined by the Penrose transform holds. For our purposes, we only use the map in one direction and in a given real slice where it can be defined in the whole hyperbolic space.} and the covariant derivative $\nabla_{\h}$ appearing in the equations of motion in the RHS arises from the Levi-Civita connection on complex hyperbolic space. On the LHS, we have the first \cech cohomology group of co-chains that (for our purposes) are represented by holomorphic functions of weight\footnote{Note that the order of the arguments of $\cO(p\,|\,q)$ is reversed with respect to the definition in \cite{tsai}.} $(-s-1-\widetilde{m},-s-1+\widetilde{m})$ on the intersection of sets in an open cover of $\MT$ and the RHS denotes the space of solutions of the equation of motion inside the braces. Strictly speaking, the LHS refers to the sheaf cohomology group that can often be approximated by the \cech one \cite{Penrose:1986ca,Huggett:1986fs}. One can also work with Dolbeault cohomology groups which is the most common modern approach \cite{Adamo:2017qyl,Woodhouse:1985id,woodhouse1976twistor}, but for the purpose of explicitly relating a cohomology class representative to a classical solution, we find the \cech representatives more useful. The relation between the Dolbeault and \cech approaches is reviewed in Appendix \ref{ap:rep}.  

For the case $s=0$ the result is
\begin{equation}\label{zerospin}
    H^1(U,\cO(-1-\widetilde{m} \,|\, -1+\widetilde{m})) \cong \left\{\nabla_{\h\,\alpha\beta}\nabla_{\h\,}^{\phantom{\h}\alpha\beta} \Psi = -2 (\widetilde{m}^2-1)\Psi\right\}\,.
\end{equation}
This is the equation of motion of a non-minimally coupled scalar. In 4d, this would be a conformally coupled scalar, but in 3d this is not the case. 

Crucially, for the cases $s=1$ and $s=2$ the relevant equations of motion above have the precise form of those arising from topologically massive Yang-Mills theory (\ref{tmymspinor}) and topologically massive gravity (\ref{tmgspinor}), respectively. More generally, these are the equations of motion of irreducible representations of the three-dimensional conformal group as seen in Sec.~\ref{sec:higherspin}. The equations coincide when a specific `real slice' of complex hyperbolic space (including the appropriate length scale $L$ by replacing the coordinates $x^\mu\xr x^\mu/L$ as described above) corresponding to the real spacetime on which the topologically massive theory is defined, is taken. The parameter $\widetilde{m}$ is then related to the topological mass parameter via $\widetilde m_{G,V}=m_{G,V}L$.\\

In the above form, the Penrose transform ensures that for each cohomology class on the LHS there is a unique solution to the spacetime equation of motion on the RHS. It is also worth noting that the Penrose transform tells us about solutions of the equations of motion locally, but not about their global properties. Additionally, since it is an isomorphism, for any solution to the spacetime equations of motion there should exist a corresponding twistor function of appropriate homogeneity that acts as a representative of a cohomology class of the minitwistor space. \\

In a more practical sense, we may construct solutions to the equations of motion, starting from elements of the \cech cohomology represented by a holomorphic function $f$ on twistor space. This is achieved using the following contour integral which is also referred to as `the Penrose transform'
\begin{equation}\label{penrosetransform}
    \Psi_{\alpha_1\ldots\alpha_{2s}}(X) = \cint\la\lambda\rd\lambda\ra\,\lambda_{\alpha_1}\ldots\lambda_{\alpha_{2s}}\,f(\mu^{\dot\alpha}_{X},\lambda_\alpha)\,,
\end{equation}
where $\la\lambda\rd\lambda\ra\equiv\lambda^{\alpha}\rd\lambda_{\alpha}=\epsilon^{\alpha\beta}\lambda_{\beta}\rd\lambda_{\alpha}$. Note that the dependence of the field on the embedding space coordinate $X$ arises entirely via $\mu^{\dot\alpha}_X$ through the incidence relations and the closed contour $\gamma$ must separate any poles of the integrand. As the incidence relation explicitly contains the coordinate $X^{\ada}$ as opposed to $x^{\ada}$, the resulting solution is defined on four-dimensional embedding space rather than the three-dimensional hyperbolic spacetime. In order to obtain the corresponding field on the three-dimensional space one must simply replace the argument $X$ with the three-dimensional coordinate $x$ from Eq.~\eqref{eq:CoordHyperbolic}.\\

It is straightforward to verify that the `integral Penrose transform' i.e. the RHS of Eq. \eqref{penrosetransform} satisfies the equation of motion on the RHS of Eq. \eqref{nonzerospin}, by applying the covariant derivative, contracting indices, and using the holomorphicity of the cohomology class representative. First, we write the expressions for the hyperbolic space covariant derivative in terms derivatives with respect to the embedding space coordinate $X^{\ada}$ (denoted $\del_{\alpha\dot\alpha}\equiv \frac{\del}{\del X^{\ada}}$) 
\begin{equation} \label{covariantderivative}
     \nabla_{\h\,\alpha\beta}f = 2 X^{\phantom{(\alpha}\dot\beta}_{(\alpha}\del_{\beta)\dot\beta}f\,,\quad \nabla_{\h\,\alpha\beta}\phi_\gamma = 2 X^{\phantom{(\alpha}\dot\beta}_{(\alpha}\del_{\beta)\dot\beta}\phi_\gamma - \delta_{(\alpha}^{\lambda}\epsilon_{\beta)\gamma}\phi_{\lambda}\,,
\end{equation}
for a function $f$ and spinor $\phi_\alpha$ \cite{tsai}.  Since $\nabla_{\h\,\alpha_1}^{\phantom{\h\alpha_1}\beta} (\lambda_{\beta}\lambda_{\alpha_2}\ldots \lambda_{\alpha_{2s}}) = (s+1) \lambda_{\alpha_1} \ldots\lambda_{\alpha_{2s}}$, we have
\begin{equation}
\begin{aligned}
    &\,\nabla_{\h\,\alpha_1}^{\phantom{\h\alpha_!} \beta} \oint_\gamma \la \lambda \rd \lambda \ra \lambda_{\beta}\lambda_{\alpha_2}\ldots \lambda_{\alpha_{2s}} f(\mu_X^{\Dot{\alpha}},\lambda_{\alpha})   \\ &\,=  \oint_\gamma \la \lambda \rd \lambda \ra \left((s+1) \lambda_{\alpha_1}  f(\mu_X^{\Dot{\alpha}},\lambda_{\alpha}) 
    -  2\lambda^\beta  X^{\phantom{(\alpha}\dot\beta}_{(\alpha_{1}}\del_{\beta)\dot\beta}f(\mu_X^{\Dot{\alpha}},\lambda_{\alpha})\right)\lambda_{\alpha_2}\ldots \lambda_{\alpha_{2s}}\,.
\end{aligned}
\end{equation}
\begin{comment}
        &\,= \oint_\gamma \la \lambda \rd \lambda \ra (s+1) \lambda_{\alpha_1} \lambda_{\alpha_2}\ldots \lambda_{\alpha_{2s}} 
    f(\mu^{\Dot{\alpha}},\lambda_{\alpha})  + \lambda_{\alpha_1} \lambda_{\alpha_2}\ldots \lambda_{\alpha_{2s}} 
    x^{\alpha_1  \Dot{\beta}} \lambda^{\beta} \frac{\partial}{\partial \mu^{\Dot{\beta}}} f(\mu^{\Dot{\alpha}},\lambda_{\alpha}) \\    
    &\,= \oint_\gamma \la \lambda \rd \lambda \ra (s+1)
     \lambda_{\alpha_1} \lambda_{\alpha_2}\ldots \lambda_{\alpha_{2s}} f(\mu^{\Dot{\alpha}},\lambda_{\alpha}) 
    +   \lambda_{\alpha_2}\ldots \lambda_{\alpha_{2s}} \mu^{  \Dot{\beta}} \lambda^{\beta} \frac{\partial}{\partial \mu^{\Dot{\beta}}} f(\mu^{\Dot{\alpha}},\lambda_{\alpha})
\end{comment}
Using anti-symmetry of the spinor inner product, the chain rule and the incidence relations one finds
\begin{equation}
    -  2\lambda^\beta  X^{\phantom{(\alpha}\dot\beta}_{(\alpha_{1}}\del_{\beta)\dot\beta}f(\mu_X^{\Dot{\alpha}},\lambda_{\alpha})=\mu_X^{\dot\beta}\lambda_{\alpha_{1}}\frac{\del }{\del\mu^{\dot\beta}}f(\mu^{\dot\alpha}_X,\lambda_\alpha)\,.
\end{equation}
Finally, Euler's homogeneous function theorem relates the derivative of a homogeneous function with its projective weight as
\begin{equation}
    \mu^{\Dot{\beta}} \frac{\partial}{\partial \mu^{\Dot{\beta}}} f(\mu^{\Dot{\alpha}},\lambda_{\alpha}) =  -(\widetilde m +s+1) f(\mu^{\Dot{\alpha}},\lambda_{\alpha}), 
\end{equation}
where the factor of $-(\widetilde m +s+1)$ is the projective weight of $f$ in $\mu^{\Dot{\alpha}}$. Hence we arrive at the final result
\begin{equation}
    \nabla_{\h\,\alpha_1}^{\phantom{\h\alpha_!} \beta} \oint_\gamma \la \lambda \rd \lambda \ra \lambda_{\beta}\lambda_{\alpha_2}\ldots \lambda_{\alpha_{2s}} f(\mu_X^{\Dot{\alpha}},\lambda_{\alpha})  =-\widetilde m \oint_\gamma \la \lambda \rd \lambda \ra  \lambda_{\alpha_1} \ldots \lambda_{\alpha_{2s}} f(\mu_X^{\Dot{\alpha}},\lambda_{\alpha})
\end{equation}
showing that the integral Penrose transform provides a solution to the equations of motion. The scalar case of Eq. \eqref{zerospin} can be shown similarly by applying the covariant derivative twice, see Appendix~\ref{ap:scalar}. This only proves one direction of the isomorphism given above, namely that each \cech  cohomology representative gives \textit{a} solution to the equations of motion via the contour integral \eqref{penrosetransform}.
%%%%%%%%%%
\subsection{\cech cohomology class representatives}
In practice, we can choose the twistor representative to be a rational function of the twistor variables up to some factor $t_0$ which has zero homogeneity (i.e. is invariant under twistor variable re-scalings). With this structure, it is easy to write representative functions with particular desired scaling weights. The Penrose transform then reads
\begin{equation}\label{penrosebracket}
    \psi_{\alpha_1\ldots\alpha_{2s}}(x)= \frac{1}{2\pi \ri}\oint_{\Gamma}\la\lambda\,\rd\lambda\ra\lambda_{\alpha_1}\ldots\lambda_{\alpha_{2s}}\frac{\prod_{i=1}^{-s-1+\widetilde m}\la\lambda A^{(i)}\ra}{\prod_{j=1}^{s+1+\widetilde m}[\mu_x B^{(j)}]}t_0(\mu_x,\lambda) \ ,
\end{equation}
where $A^{(i)}_\alpha$ and $B^{(j)}_{\dot{\alpha}}$ are arbitrary constant spinors. The choice of these constant spinors will contribute to the pole structure of the representative function. Also note that the field $\psi_{\alpha_1\ldots\alpha_{2s}}(x)$ is a field on hyperbolic space not on embedding space, so the incidence relation in Eq.~\eqref{eq:incidence} is used with $X \rightarrow x$, where the hyperbolic space coordinate is given as in Eq.~\eqref{eq:CoordHyperbolic}. We will  additionally focus on representatives that factorise as
\begin{equation} \label{eq:factorize}
f_{s, \widetilde{m}}(\mu^{\dot\alpha},\lambda_\alpha)=g_{s, \widetilde{m}}(\mu^{\dot\alpha}) h_{s, \widetilde{m}}(\lambda_\alpha) \ .
\end{equation}
As a simple example, from the Penrose transform we can obtain a scalar field with $\widetilde m =0$ to set the notation for the rest of the paper. Plugging $t_0=1$, $s=0$, and $\widetilde m =0$ in Eq. \eqref{penrosebracket}, we have 
\begin{equation}\label{scalarm0}
    \begin{aligned}
        \phi(x)&\,=\frac{1}{2\pi\ri}\oint_\gamma \la\lambda \rd\lambda\ra\frac{1}{\la\lambda A\ra[\mu_x B]}\, \\
        &\,= -\frac{1}{2\pi\ri}\oint_\gamma \rd \xi\,\frac{1}{(\xi A_0-A_1)(\xi B_{0}(x) - B_{1}(x))}\, 
    \end{aligned}
\end{equation}
where we have locally parametrised $\lambda_{\alpha}=(1, \xi)$.
The $x$-dependent un-dotted spinor, $B_{\alpha}(x)$ is defined from
from the constant \textit{dotted} spinor ${B}_{\dot\alpha}$ via
\begin{equation} \label{eq:spinorX}
    {B}(x)_\alpha\equiv-x_{\alpha}^{\phantom{\alpha}\dot\alpha}{B}_{\dot\alpha}\,.
\end{equation}
From now on, wherever an un-dotted spinor carries an argument $(x)$ it should be understood that it is defined in this way using a constant dotted spinor. Note that we can choose the arbitrary spinors so that the contour encloses the only pole at $\xi=A_1/A_0$ then
\begin{equation}\label{scalarm0sol}
\phi(x)=-\frac{1}{\la A \,B(x)\ra}=-\frac{1}{A_\alpha x^{\alpha\dot\alpha}B_{\dot\alpha}}= -\frac{1}{2v\cdot x}\,,
\end{equation}
which indeed satisfies the equation of motion Eq. \eqref{zerospin}. In the last equality, we have defined a constant null 4-vector $v_{\alpha\dot{\alpha}}$ from the bi-spinor $A_\alpha B_{\dot\alpha}$ as simply, $v_{\alpha\dot{\alpha}}=A_\alpha B_{\dot\alpha}$.\\
Finally, we note that not all representatives are of the form of Eq. \eqref{eq:factorize}. For instance, taking $a^{\alpha}_{\dot{\alpha}}$ to be a constant bispinor,
\begin{equation}
    f_{s=0, \tilde{m}=2}(\mu^{\dot\alpha},\lambda_\alpha)= \frac{a^{\alpha}_{\dot{\alpha}} \lambda_\alpha \mu^{\dot{\alpha}}}{[\mu A]^4} \ ,
\end{equation}
does not factorise and non-elementary representatives of this type will not be examined.

%%%%%%%%%%%%%%%%%%%%
\subsection{Real hyperbolic space}
As already mentioned, we have been working so far with complex coordinates where the signature does not have any meaning. The different real signatures (for example: Euclidean, Lorentzian, and split) can be obtained by selecting a reality condition. This is achieved by defining a conjugation and requiring $\hat{x}^{\alpha\dot{\alpha}}=x^{\alpha\dot{\alpha}}$, from which the conjugation on $\lambda_{\alpha}$ and $\mu^{\dot{\alpha}}$ can be deduced, as we now show explicitly. Three-dimensional real hyperbolic space can be embedded in four-dimensional Minkowski spacetime. We must therefore first pick the reality condition in four dimensions that gives rise to a Lorentzian signature. This is done by requiring that the coordinates are real and hence
\begin{equation}
    \hat{X}^{\alpha\dot{\alpha}} = (X^{\alpha\dot{\alpha}})^{\dagger} \ .
    \label{eq:conjugation}
\end{equation}
A useful coordinate choice for hyperbolic space is the Poincaré coordinates which can be obtained by picking the embedding
\begin{equation}\label{poincareemb}
    X^{\alpha\dot{\alpha}} = L \begin{pmatrix}
        1 && y_1 - \ri y_2 \\
        y_1 + \ri y_2 && y_0^2+y_1^2+y_2^2
    \end{pmatrix},
\end{equation}
where $y_i$ are now real dimensionless coordinates. This gives rise to the 3d line element
\begin{equation}\label{euclideanadspoincare}
    ds^2= -\frac{L^2}{y_0^2}\left(dy_0^2+dy_1^2+dy_2^2\right) ,
\end{equation}
as can be checked by plugging these coordinates in Eq. \eqref{embedded}.
These coordinates cover the whole space, contrary to Lorentzian signature where they cover only half of the spacetime. One can see that, $y_0\xrightarrow[]{} 0$ corresponds to the boundary, while $y_0\xrightarrow[]{} \infty$ corresponds to a point deep in the bulk.

The conjugation chosen in Eq.~\eqref{eq:conjugation} induces a conjugation on spinors in which positive and negative chirality spinors are exchanged, that is, $\alpha$ and $\dot{\alpha}$ are exchanged under conjugation. Thus, under conjugation we have 
\begin{equation}
\begin{aligned}
    \lambda_{\alpha} = \begin{pmatrix}
        a \\
        b
    \end{pmatrix} \xrightarrow[]{} \bar{\lambda}_{\dot{\alpha}} = \begin{pmatrix}
        a^* \\
        b^*
    \end{pmatrix} \ , \\
    \mu_{\dot{\alpha}} = \begin{pmatrix}
        c \\
        d
    \end{pmatrix} \xrightarrow[]{} \bar{\mu}_{\alpha} = \begin{pmatrix}
        c^* \\
        d^*
    \end{pmatrix} \ .
\end{aligned}
\end{equation}
In this real slicing, a point in minitwistor space corresponds to the real slicing of the null geodesic planes, which gives the geodesics of the real hyperbolic space. Such geodesics are described by
\begin{equation} \label{eq:geod}
    X^{\alpha \dot{\alpha}}(t) = w^{\alpha\dot{\alpha}} + t z^{\alpha\dot{\alpha}} \ ,
\end{equation}
where $t=1/\braket{\bar{\mu} \lambda}$ is a parameter whose variation gives the coordinates of the geodesic in the bulk and $w^{\alpha\dot{\alpha}}$ and $z^{\alpha\dot{\alpha}}$ are two points in the boundary, which is an $S^2$. These two boundary points are
\begin{equation}
\begin{aligned}
    z^{\alpha\dot{\alpha}} = \bar{\mu}^{\alpha}\mu^{\dot{\alpha}}, \\
    w^{\alpha\dot{\alpha}} = \lambda^{\alpha}\bar{\lambda}^{\dot{\alpha}}.
\end{aligned}
\end{equation}
Note that Eq.~\eqref{eq:geod} implies the incidence relation in Eq.~\eqref{eq:incidence}. The two boundary points are distinct unless $\bar{\mu}^{\alpha} \propto \lambda^{\alpha}$. The correspondence is therefore well defined unless $\braket{\bar{\mu} \lambda}=0$ (the so-called anti-holomorphic diagonal), which prompts the following definition for the minitwistor space of EAdS that we will be working on
\begin{equation}
    \mathbb{MT}=\left\{(\mu^{\dot\alpha},\lambda_\alpha)\in\cp^1\times\cp^1\,|\,\la\Bar{\mu}\lambda\ra\neq0\right\}\cong\{\cp^1\times\cp^1\}\setminus\cp^1\,.
\end{equation}
If instead, we had started with split signature in embedding space, we would have obtained the minitwistor space of AdS$_3/\mathbb{Z}$ which is given by $\mathbb{RP}^1\times\mathbb{RP}^1$.

\section{Double copy prescription in hyperbolic spaces} \label{sec:DC}
After briefly reviewing topologically massive theories and how solutions to their equations of motion can arise from minitwistor space via the Penrose transform, we proceed to introduce our prescription for the double copy in hyperbolic backgrounds. The classical double copy relates gravitational solutions ($s=2$) with two spin one solutions divided by a scalar. In position space, the double copy takes the form of
\begin{equation}
    C_{\alpha\beta\gamma\delta} =\frac{f_{(\alpha \beta}f_{\gamma \delta)}}{\phi}, \label{eq:CottonDCposition}
\end{equation}
where as in Section 2, $C_{\alpha\beta\gamma\delta}$ is the Cotton spinor of a TMG solution, $f_{\alpha \beta}$ is the dual field strength of a linearised TMYM field, and $\phi$ is a scalar satisfying the equation of motion in the RHS of Eq.~\eqref{zerospin}. This is the Cotton double copy from \cite{CarrilloGonzalez:2022mxx,Emond:2022uaf} and it is known to hold for type N solutions in maximally symmetric backgrounds and has been shown to fail for some linearised type D solutions in asymptotically flat spacetimes \cite{CarrilloGonzalez:2022ggn}. Note that in Eq.~\eqref{eq:CottonDCposition} the LHS and RHS do not have the same mass dimensions so in principle the Cotton spinor is only proportional to the RHS of of Eq.~\eqref{eq:CottonDCposition} up to a dimensionful constant factor. While this depends on the choice of normalisation of the fields, in flat space, we only have one dimensionful scale given by the mass of the particles which is the same for all spins, so there is no ambiguity in this relation. On the other hand, in AdS we have four scales. These are the AdS length and the mass of the graviton, gauge field, and scalar which, as we will see below, can all be different in the double copy relation.  The combination of these scales could give rise to additional numerical factors on the RHS. We will make further comments on this issue and its possible consequences in the conclusion.

Here, we will focus on the double copy in twistor space for hyperbolic spacetimes and we will also analyse whether the position double copy holds for special solutions. In twistor space, the double copy relation is realised between \cech cohomology class representatives. The fact that the relation holds is a simple consequence of the homogeneity of the twistor representatives (see Eqs. \eqref{scaling}, \eqref{nonzerospin}, \eqref{zerospin}). One can define a spin two representative as
\begin{equation}
f_{s=2, \widetilde m_G}( \mu^{\dot\alpha},\lambda_{\alpha})
=\frac{f_{s=1, \widetilde m_V}(\mu^{\dot\alpha},\lambda_\alpha) f'_{s=1, \widetilde m_V'}(\mu^{\dot\alpha},\lambda_\alpha)}{f_{s=0, \widetilde m_\phi}( \mu^{\dot\alpha},\lambda_\alpha)} \ ,
\label{eq:DCtwistors}
\end{equation}
where the mass $m_G$ of the spin two field is related to the masses $m_V$ and $m_\phi$ of the spin one fields and spin zero field respectively as
\begin{equation}
\widetilde m_G=\widetilde m_V+\widetilde m_V'-\widetilde m_\phi \ .
\end{equation}
It is easy to see that this representative has the correct homogeneity  
\begin{align}
f_{s=2, \widetilde m_G}(u \mu^{\dot\alpha},v\lambda_{\alpha})&=
    v^{\widetilde m_V+\widetilde m_V'-\widetilde m_\phi-3} u^{ -(\widetilde m_V+\widetilde m_V'-\widetilde m_\phi)-3} f_{s=2, \widetilde m_V+\widetilde m_V'-\widetilde m_\phi}(\mu^{\dot\alpha},\lambda_\alpha) \ .
\end{align}
In the following, we proceed by considering symmetric double copies, where both spin one fields have the same mass, although in Section~\ref{sec:beyondtypesND}, we will consider asymmetric double copies. In the symmetric case, the twistor double copy requires
\begin{equation}\label{masses}
    \widetilde m_{G}=  2  \widetilde m_{V}- \widetilde m_{\phi} \ .
\end{equation}
When constructing the topologically massive double copy in flat backgrounds, one fixes the mass of the gluons to be the same as the mass of the gravitons. This is consistent with the BCJ double copy for scattering amplitudes where the denominator that arises from propagators is kept fixed. Choosing this prescription in the present case would lead to
\begin{equation}\label{wrong}
     \widetilde m_{G}= \widetilde m_{V}=\widetilde m_{\phi} \quad \text{(incorrect prescription)}.
\end{equation}
However, we will see that, while this prescription can give rise to a seemingly correct double copy of wave solutions \cite{CarrilloGonzalez:2022mxx}, it is not the correct prescription in general. We will argue that the correct prescription is rather
\begin{equation}\label{correctMass}
\boxed{\widetilde m_G=\widetilde m_V+\text{sign}(\widetilde m_{G})=\widetilde m_\phi + 2 \ \text{sign}(\widetilde m_{G})\ ,}
\end{equation}
or equivalently
\begin{equation}\label{correct}
\Delta_{\pm,G}= \Delta_{\pm,V}\pm \text{sign}(\eta_{G})= \Delta_{\pm,\phi}\pm 2 \  \text{sign}(\eta_{G})\ .
\end{equation}

We can additionally choose \cech cohomology class representatives whose $\mu^{\dot\alpha}$ and $\lambda_\alpha$ dependence factorises
as in Eq.~\eqref{eq:factorize}. This choice of representatives will highly simplify the double copy when considering the prescription in Eq.~\eqref{correct}. Eq.~\eqref{correct} implies that the representatives for all $s=0,1,2$ have the same scaling weight with respect to $\lambda_\alpha$. A simple example of this is
\begin{equation}
    h_{s=2, \widetilde{m}_{\phi}+2}(\lambda_\alpha)=h_{s=1, \widetilde{m}_{\phi}+1}(\lambda_\alpha)=h'_{s=1,\widetilde{m}_{\phi}+1}(\lambda_\alpha)=h_{s=0,\widetilde{m}_{\phi}}(\lambda_\alpha)= \braket{\lambda A}^{\widetilde{m}_{\phi}-1},
\end{equation}
where $\widetilde{m}$ is the mass of the scalar and $A_\alpha$ a fixed constant spinor. In this case, the $\lambda$ dependence drops from the twistor double copy and one can simply express the double copy as 
\begin{equation}\label{simplification}
    g_{s=2, \widetilde{m}_\phi+2}(\mu^{\dot\alpha})=\frac{g_{s=1, \widetilde{m}_\phi+1}(\mu^{\dot\alpha}) g'_{s=1, \widetilde{m}_\phi+1}(\mu^{\dot\alpha})}{g_{s=0, \widetilde{m}_\phi}(\mu^{\dot\alpha})}.
\end{equation}
The same is true for the opposite helicity by swapping $\lambda$ and $\mu$. In the following sections, we will show how these simplifications arise in the case of a double copy for Petrov type N and type D gravitational solutions. A short review of the Petrov classification of TMG solutions is given in Appendix~\ref{app:petrov}. These cases are particularly interesting since they correspond to exact solutions to TMG, not just linearised ones.

%%%%%%%%
\subsection{Consequences of the double copy on the dual theory}
We will see in Section~\ref{sectiontyped} that this prescription leads to a physical double copy of black hole solutions. Additionally, in the context of AdS/CFT, the prescription in Eq. \eqref{correctMass} double copies bulk fields dual to conserved currents, to bulk fields dual to conserved currents, contrary to the incorrect prescription in Eq. \eqref{wrong}.  In three dimensions, a spinning bulk field is dual to a conserved current in the boundary CFT when $\Delta_{+,s}=s$, that is, when the unitarity bound in Eq.~\eqref{unitaritybound} is saturated. Note that for the spin one case, this corresponds to a massless gauge field and for the spin two it gives the chiral point of TMG \cite{Li:2008dq,Li:2008yz,Maloney:2009ck,Giribet:2008bw}. For a scalar, the choice $\Delta_{+,0}=0$ saturates the Breitenlohner-Freedman bound \cite{Breitenlohner:1982bm,Breitenlohner:1982jf} and gives the scalar a vanishing effective mass so that its equation of motion is $\nabla^2\phi=0$. Eq.~\eqref{correctMass} also keeps the denominator of bulk-to-boundary propagators in coordinate space fixed \cite{Costa:2011mg, Roehrig:2020kck}, which hints to a possible realisation of a double copy for correlators in coordinate space. We will see this explicitly in Section \ref{sec:typeN}.

Another interesting consequence of our prescription can be seen as follows. As shown explicitly in \cite{Carlip:2008jk}, the linearised actions of the perturbations of topologically massive Maxwell and topologically massive gravity around AdS can be written through field redefinitions as the action of a scalar propagating in AdS, which can be understood from the fact that in 2+1 dimensions spin is a pseudo-scalar. This simplification is easily seen when working in lightcone coordinates defined as $y_{\pm}= \frac{y_1\pm y_2}{\sqrt{2}}$. Additionally, we label the components of the fields as $A_0\equiv A_{y_0}$, $A_{\pm} \equiv A_{y_{\pm}}$ such that $A_{\mu}=(A_+,A_-,A_0)$. In AdS, the field strength tensor components $(F_{+z},F_{+-},F_{-0})$ of topologically massive Maxwell satisfy scalar field equations with different mass parameters. A similar feature arises for the linearised curvature tensor in the spin 2 case.  Explicitly, we have the following scalar equations of motions in lightcone coordinates
\begin{equation}\label{components eq}
\begin{aligned}\left(2\partial_-\partial_++\partial_0^2-\frac{\widetilde m_{\phi}^2-1/4}{y_0^2}\right)\frac{\phi}{\sqrt{y_0}}&\,=0,\\
\left(2\partial_-\partial_++\partial_0^2-\frac{(\widetilde m_{V}-1)^2-1/4}{y_0^2}\right)\sqrt{y_0}F_{+0}&\,=0,\\
\left(2\partial_-\partial_++\partial_0^2-\frac{(\widetilde m_{G}-2)^2-1/4}{y_0^2}\right)\frac{\mathcal{H}_{++}}{\sqrt{y_0}}&\,=0,\\
\end{aligned}
\end{equation}
and similarly for the $-$ components. The solutions to these equations are well known and are given in terms of Bessel functions. Here $\mathcal{H}_{\mu \nu} \equiv (G_{\mu\nu}+g_{\mu\nu})_{\text{linear}}$ is the linearised Cotton tensor (up to a factor of $\widetilde m_G$). The above equations follow directly from the equations of motion obeyed by the gauge fields and linearised graviton respectively. It is interesting to note that matching the third terms in Eq. \eqref{components eq} leads then to the prescription of Eq. \eqref{correctMass} for $\widetilde{m}_G>0$. The masses and asymptotics of the relevant curvature components are summarised in the following table.

\begin{center}
\begin{tabular}{||c c c c c||} 
 \hline
 $s=0$ & $s=1$ & $s=2$ & $\widetilde m_{eff}^2$ & Asymptotics $y_0 \xrightarrow{} 0$\\ [0.5ex] 
 \hline\hline
  &  & $\mathcal{H}_{++}$ & $(\widetilde m_{G}-2)^2-1$ & $y_0^{\Delta_G-2}$\\ 
 \hline
  & $F_{+0}$ & \phantom{$\mathcal{H}_{+z}$} & $(\widetilde m_{V}-1)^2-1$ & $y_0^{\Delta_V-2}$\\
 \hline
 $\phi$ & \phantom{$F_{+-}$} & \phantom{$\mathcal{H}_{+-}$} & $\widetilde m_{\phi}^2-1$ & $y_0^{\Delta_\phi}$\\
 \hline
  & $F_{-0}$ & \phantom{$\mathcal{H}_{-z}$} & $(\widetilde m_{V}+1)^2-1$ & $y_0^{\Delta_V}$ \\
 \hline
  &  & $\mathcal{H}_{--}$ & $(\widetilde m_{G}+2)^2-1$ & $y_0^{\Delta_G+2}$\\ [0.1ex] 
 \hline
\end{tabular}
\end{center}
The asymptotics refer to the scaling of the relevant field components near the Poincaré boundary with respect to the radial coordinate $y_0$.  From this, we see that $\mathcal{H}_{++}$ and $F_{+0}$ are the leading components at the boundary. In the AdS/CFT context, the sources / dual currents at the boundary are given by \cite{Klebanov:1999tb, Kabat:2012hp, Sundrum:2011ic}
\begin{equation}
    \begin{aligned}
        \mathcal{O}(x)&\,=\lim_{y_0\to 0}\frac{\phi(y_0,x)}{y_0^{\Delta_\phi}}\ ,\\
        J_{\mu}(x)&\,=\lim_{y_0\to 0}\frac{A_{\mu}(y_0,x)}{y_0^{\Delta_V-1}}=\frac{1}{\Delta_V-1}\lim_{y_0\to 0}\frac{F_{0\mu}(y_0,x)}{y_0^{\Delta_V-2}}\ ,\\
        O_{\mu \nu}(x)&\,=\lim_{y_0\to 0}\frac{h_{\mu \nu}(y_0,x)}{y_0^{\Delta_G-2}}=\frac{2}{\Delta_G(\Delta_G-2)}\lim_{y_0\to 0}\frac{\mathcal{H}_{\mu \nu}(y_0,x)}{y_0^{\Delta_G-2}} \ ,
    \end{aligned}
\end{equation}
where $\mu=\pm$. Therefore, the components $\mathcal{H}_{++}$ and $F_{+0}$ give the CFT currents. Now, we can see that requiring that near the boundary the following have the same scaling
\begin{equation}
    \phi(y_0,x)\sim A_{+}(y_0,x) \sim  h_{++}(y_0,x)\sim y_0^{\Delta_\phi}
\end{equation}
requires that the masses of $\phi$, $\mathcal{H}_{++}$, and $F_{+0}$ in Eq.\eqref{components eq} are equal which is satisfied in our double copy prescription. This also tells us that near the boundary, our double copy relates the dual CFT currents.

In the above, we have considered boundary conditions that lead to the standard quantisation. Instead, we could have considered the alternate quantisation in which the fall-off of the wavefunction goes like $y_0^{\Delta_-}$ instead of $y_0^{\Delta_+}$ which is allowed for small masses \cite{Klebanov:1999tb}. In that case, the CFT currents would be given by the components $F_{-0}$ and $\mathcal{H}_{--}$ and the double copy would correspond to Eq. \eqref{correctMass} for $\widetilde{m}_G<0$.

%%%%%%%%%%%
\subsection{Other double copies}
In this subsection, we look at the possibility of generalising the double copy prescription in Eq.~\eqref{correct}. We will consider higher-spin particles and a double copy between different particles with different helicities.

\paragraph{Higher Spins:}
Double copy relations for higher spin particles in $d>3$ have been considered in \cite{Didenko:2009td,Ponomarev:2017nrr,Monteiro:2022lwm,Monteiro:2022xwq,Didenko:2022qxq,Didenko:2021vdb}. In analogy, we construct a three-dimensional version in this subsection. Besides the standard double copy relation for gravity as the square of a gauge theory that we have described above, we can see that a higher-spin version of the double copy also exists as
\begin{equation}
f_{2s, \widetilde m_{2s}}( \mu^{\dot\alpha},\lambda_{\alpha})
=\frac{f_{s, \widetilde m_s}(\mu^{\dot\alpha},\lambda_\alpha) f'_{s, \widetilde m_s'}(\mu^{\dot\alpha},\lambda_\alpha)}{f_{s=0, \widetilde m_\phi}( \mu^{\dot\alpha},\lambda_\alpha)} \ ,
\label{tdchigherspin}
\end{equation}
where we have 
\begin{equation}\label{masseshigherspin}
   \widetilde m_{2s}=    \widetilde m_{s}+\widetilde m_{s'}- \widetilde m_{\phi} \ .
\end{equation}
As before this does not completely fix the conformal dimension (or equivalently the mass). However,  generalising Eq.~\eqref{correct} we can take 
\begin{equation}\label{eq:DeltaSpin}
\boxed{ \widetilde m_{s}= \widetilde m_{\phi}+ \text{sign}(\widetilde m_{2s})s \ .}
\end{equation}
This prescription generalises to higher-spins the properties described in the previous section.
\paragraph{Mixed Helicities:}
One can write a double copy between particles with opposite helicity as
\begin{equation}
    \frac{f_{\widetilde{m}_{+,s}}( \mu,\lambda)f_{\widetilde{m}_{-,s}}(\mu,\lambda)}{f_{\widetilde{m}_{\phi}}(\mu,\lambda)}= f_{\widetilde{m}_{2s}}(\mu,\lambda),
\end{equation}
where $\widetilde m_{\pm,s}= \pm(\widetilde{m}_\phi+s)$. To obtain the correct homogeneity of the representatives
we need to set $\widetilde{m}_{2s}=-\widetilde{m}_{\phi}$. In general, this violates our prescription in Eq.~\eqref{eq:DeltaSpin} except at the special point $\widetilde{m}_{2s}=\pm s$. At this point, the mass of the spin $s$ field that is being double copied vanishes, so we can no longer interpret this as a double copy between particles of opposite helicities. In fact, for the usual double copy of gauge fields with $s=1$, this corresponds to the double copy of fields dual to conserved currents that was discussed in the previous subsection. Thus, to have a double copy of mixed helicities we need to abandon the prescription of Eq.~\eqref{eq:DeltaSpin} which we will not consider in this paper.

%%%%%%%%%%%
\subsection{Ambiguity in the double copy}
We conclude this section by highlighting the fact that there exists a cohomological ambiguity in the choice of twistor representative that enters the twistor double copy. Since the Penrose transform gives an isomorphism between cohomology classes and solutions to equations of motion, any representative will give the same solution, but not any representative satisfies a double copy relation. One can always add a holomorphic function to $f$ in the Penrose transform of Eq. \eqref{penrosetransform} without changing the position space field and it is not a priori obvious which representative should be picked. Furthermore, given a set of representatives satisfying the double copy, a generic change of representative by adding a holomorphic function will not satisfy the double copy.

In four dimensions with Euclidean signature, one prescription to choose the representatives that satisfy the double copy is to work with Dolbeault cohomology (instead of Čech cohomology) and take the harmonic representative, which is unique \cite{Chacon:2021lox}. Another solution to this issue was given for radiative spacetimes, where the characteristic data at null infinity can be used to uniquely determine a specific Dolbeault twistor representative \cite{mason:1986, Adamo:2021dfg}. A third prescription was given in \cite{Luna:2022dxo}, where it was shown how certain twistor representatives can be obtained directly from scattering amplitudes by taking the classical limit similar to earlier approaches in \cite{Kosower:2018adc,Monteiro:2020plf,Monteiro:2021ztt,Guevara:2021yud}. In our case, we leave it as an open question for future work to understand whether the representatives that we find can be related to correlation functions. 
 
%%%%%%%%%%%%%%%%%%%%
\section{Type N double copy in topologically massive gravity} \label{sec:typeN}
In this section, we are going to show that Petrov type N solutions in TMG can be written as a double copy of the analogue wave\footnote{We refer to the solutions as wave solutions even if we work on Euclidean signature, since one can straightforwardly switch to the standard Lorentzian signature wave solutions by performing an analytic continuation of the time coordinate.} solutions of Topologically Massive Gauge Theories. Earlier work showing this in coordinate space by using the flat space double copy prescription can be found in \cite{CarrilloGonzalez:2022mxx}. Since type N solutions can be written in Kerr-Schild form, the linearised solutions that we find from twistor space will be exact solutions of TMG. Hence, we start by reviewing the Kerr-Schild construction of type N solutions and showing the coordinate space Kerr-Schild double copy. Then we write explicit representatives that give rise to wave solutions for generic spin $s$ fields and satisfy a double copy relation. 
%%%%%%%%%%%%%%%%
\subsection{Kerr-Schild construction and double copy}
Gravitational solutions of Petrov type II can have their metric expressed in the so-called Kerr-Schild form that linearises Einstein's equations in General Relativity. The Kerr-Schild form of a metric $g_{\mu\nu}$ can generally be written as
\begin{equation}
    g_{\mu\nu}=\bar{g}_{\mu\nu}+\phi k_\mu k_\nu,
    \label{eq:KSmetric}
\end{equation}
where $\bar{g}_{\mu\nu}$ is the base metric, $\phi$ is a scalar field that depends on the coordinates of the metric, and $k_\mu$ is the null and geodetic \textit{Kerr-Schild (co-)vector}, 
\begin{equation}
    g^{\mu\nu}k_\mu k_\nu=\bar{g}^{\mu\nu}k_\mu k_\nu=0,\quad\quad k^\mu\nabla_\mu k^\nu=k^\mu \bar{\nabla}_\mu k^\nu=0,
    \label{eq:ksvec1}
\end{equation}
where we have placed (and continue to, hereafter) bars above objects related to the background metric.  

In this section, we will focus on solutions of TMG that admit a null Killing vector, these are the pp-waves \cite{Chow:2009km,Ayon-Beato:2004nrg,Ayon-Beato:2005gdo,Olmez:2005by}. The Kerr-Schild form of pp-wave solutions is given by
\begin{equation}
      ds^2= -\frac{L^2}{y_0^2}\left(dy_0^2+dudv +y_0^{\widetilde m_G+1}f(u)du^2 \right) , \label{eq:ppwave}
\end{equation}
where $u=y_1+iy_2$ and $v=y_1-iy_2$, that is, our coordinates are $x^\mu=(u,v,y_0)^\mu$. Here we consider the negative chirality wave solution, but the other chirality can be obtained analogously. We remain agnostic of the value of the mass which defines whether the solution is asymptotically AdS or not \cite{Henneaux:2009pw,Henneaux:2010fy}.  Note that we can recover the real null coordinates by analytically continuing $y_2\rightarrow i y_2$ and going to Lorentzian signature. We can see that the base metric is given by AdS in Poincare coordinates and 
\begin{equation}\label{eq:KSvectANDphi}
k_\mu d x^\mu = du \ , \quad \phi= L^{2}y_0^{\widetilde m_G-1}f(u) \ .
\end{equation}
The Cotton tensor for this solution is
\begin{equation} \label{eq:CottonPPwave}
    C^{\mu\nu}= \left(
\begin{array}{ccc}
 0 & 0 & 0 \\
 0 & \frac{2 f(u) \widetilde\mu_G \left(\widetilde m_G^2-1\right) y_0^{\widetilde m_G+3}}{L^4}
   & 0 \\
 0 & 0 & 0 \\
\end{array}
\right)^{\mu\nu} \ .
\end{equation}
\paragraph{Single Copy}
To find the single copy, we use the Kerr-Schild ansatz for the gauge field
\begin{equation} \label{eq:KSsinglecopy}
     A_\mu^a = \phi k_\mu c^a \ ,
\end{equation}
where $c^a$ is a constant colour charge factor in the adjoint. As mentioned in Section \ref{sec:TMYM}, this linearises the equations of motion such that we are effectively working with Maxwell-Chern-Simons solutions and the equations of motion are now
\begin{equation}
    \bar{\nabla}_\mu F^{\mu\nu}-\frac{m_V}{2}\bar{\epsilon}^{\nu\alpha\beta}F_{\alpha\beta} = j^\nu \ .
    \label{eq:MCSEOM}
\end{equation}
The field strength is given by
\begin{equation}
  F^{\mu\nu}=  \left(
\begin{array}{ccc}
 0 & 0 & 0 \\
 0 & 0 & -\frac{2 f(u) \left(\widetilde m_G-1\right) y_0^{\widetilde m_G+2}}{L^2} \\
 0 & \frac{2 f(u) \left(\widetilde m_G-1\right) y_0^{\widetilde m_G+2}}{L^2} & 0 \\
\end{array}
\right)^{\mu\nu}
    \label{eq:typeNsinglecopyF}
\end{equation} 
and the source, as a result of the choices above, takes the form
\begin{equation}
  j^\nu=  \left(
\begin{array}{c}
 0 \\
 \frac{2 f(u) \left(\widetilde m_G-1\right) y_0^{\widetilde m_G+1} \left(\widetilde m_G-\widetilde m
   _V-1\right)}{L^2} \\
 0 \\
\end{array}
\right)^\nu. 
    \label{eq:typeNsinglecopysrc}
\end{equation} 
From this we can see that the source vanishes when
\begin{equation}
   \widetilde m_V=\widetilde m_G-1\ . 
\end{equation}
\paragraph{Zeroth Copy}
The zeroth copy corresponds to a linearised biadjoint scalar theory for a scalar field given by
\begin{equation}
    \Phi^{a\tilde a}=c^a c^{\tilde{a}}\Phi \ ,
\end{equation}
where $c^a$ and $c^{\tilde a}$ are constant colour charge factors in the adjoint of $G$ and $\tilde{G}$ respectively. The linearised equations of motion for a biadjoint scalar in a AdS background are 
\begin{equation}
    \bar{\nabla}^2\Phi +\Big(\widetilde m_{\phi}^2-1\Big)\Phi = \mathcal{J} \ ,
    \label{eq:singlecopyeq}
\end{equation}
where we have used $R=-6 \bar{\Lambda}=6/\bar{L}$ As before, identifying $\Phi\equiv\phi$, the source is of the form 
\begin{equation}
    \mathcal{J} = f(u) y_0^{m_G-1} \left(\widetilde m_\phi^2-\left(\widetilde m_G-2\right){}^2\right).
\end{equation}
In this case, we see that the source vanishes when
\begin{equation}
    \widetilde m _\phi = \pm(\widetilde m_G-2).
\end{equation}
Note that the values of the masses for the zeroth (when choosing the $+$ sign) and single copies that give the expected vacuum wave solutions correspond to the double copy prescription in Eq.~\eqref{correctMass} for positive helicity. If we had chosen the opposite chirality of the pp-wave solution, we would have instead obtained the relation for the masses with negative helicity.
%%%%%%%%
\subsection{Type N twistor double copy}
In this section, we are going to specialise to wave solutions in which the spin $s$ fields are of the form
\begin{equation}\label{fieldtypeN}
    \Psi_{\alpha_1\ldots\alpha_{2s}}(x)=\Psi(x) A_{\alpha_1}\ldots A_{\alpha_{2s}} \ .
\end{equation}
As noted in \cite{Penrose:1985bww, White:2020sfn, Chacon:2021wbr} and explained in Appendix \ref{app:petrov}, a spin $s$ spacetime field produced by the integral Penrose transform will have a principal spinor of multiplicity at least $(2s+1-J)$ if it has a single order $J$ pole inside the contour. 
Fields of the form of Eq. \eqref{fieldtypeN} can then be obtained via the Penrose transform by considering cohomology class representatives that have a simple pole enclosed inside the contour. In the gravitational context, these correspond to Petrov type N solutions. 

For the non-zero spin fields, a straightforward generalisation of Eq.~\eqref{scalarm0} gives the desired wave solutions
\begin{equation}\label{bulktoboundary}
\begin{aligned}
        \Psi_{\alpha_1\ldots\alpha_{2s}}(x)&=\cint \la\lambda\rd\lambda\ra\lambda_{\alpha_1}\ldots\lambda_{\alpha_{2s}}\frac{1}{\la\lambda A\ra}\la\lambda B\ra^{\widetilde m-s}[\mu_x C]^{-s-1-\widetilde m}\\
        &=-A_{\alpha_1}\ldots A_{\alpha_{2s}}\la A B\ra^{\widetilde m-s}(A_{\alpha} x^{\alpha\dot\alpha} C_{\dot\alpha})^{-s-1-\widetilde m}\\
        &=-\la A B\ra^{\widetilde m-s} \ \frac{ A_{\alpha_1}\ldots A_{\alpha_{2s}}}{(2v\cdot x)^{s+1+\widetilde m}}\,,
\end{aligned}
\end{equation}
where we assumed that the contour only encloses the simple pole at $\xi=A_1/A_0$ and we defined the null vector $v_{\alpha\dot\alpha}= A_\alpha C_{\dot\alpha}$.  These solutions are the analogues of plane waves in flat space and correspond, up to normalisation, to bulk-to-boundary propagators in AdS \cite{Costa:2014kfa,Kutasov:1999xu,deBoer:1998gyt,Roehrig:2020kck}
\begin{equation}\label{Deltawave}
    \Psi_{\alpha_1\ldots\alpha_{2s}}(x)\propto \frac{ A_{\alpha_1}\ldots A_{\alpha_{2s}}}{(v\cdot x)^{s+\Delta_+}} \ ,
\end{equation}
where the polarisations are given by $A_\alpha A_\beta$. This computation was explicitly shown in \cite{Bu:2023cef} for the scalar case.  Had we integrated around a simple pole in $[\mu_x D]$ instead (where $D_{\dot\alpha}$ is constant), the result would have been identical with the exception that the sign of the mass would have been reversed. Equivalently, $\Delta_+$ would have been replaced by $\Delta_-$ in Eq. \eqref{Deltawave}. More explicitly, the bulk-to-boundary propagator would now be
\begin{equation}
    \begin{aligned}
        \Psi_{\alpha_1\ldots\alpha_{2s}}(x)&=\cint \la\lambda\rd\lambda\ra\lambda_{\alpha_1}\ldots\lambda_{\alpha_{2s}}\frac{1}{[\mu_x D]}\la\lambda B\ra^{\tilde m-s-1}[\mu_x C]^{-s-\tilde m}\\
        &= -[DC]^{-\tilde m-s}\  \frac{D_{\alpha_1}(x)\ldots D_{\alpha_{2s}}(x)}{\left(2w\cdot x\right)^{s+\Delta_-}}\,,
\end{aligned}
\end{equation}
where we have defined $w_{\alpha\dot\alpha}\equiv - B_\alpha  D_{\dot\alpha}$ above and $D_{\alpha}(x)$ are defined as in Eq.~\eqref{eq:spinorX}. Note that these representatives and coordinate space solutions can be obtained from a dimensional reduction of plane waves in 4d, see Appendix \ref{app:dimred}. One can easily see that these fields satisfy the double copy in twistor space both with the equal mass prescription in Eq.~\eqref{wrong} and with our proposed prescription in Eq.~\eqref{correct}. Additionally, these fields satisfy the Cotton double copy from Eq.~\eqref{eq:CottonDCposition} in position space for both prescriptions. This was previously proved for all type N solutions in maximally symmetric spacetimes in \cite{CarrilloGonzalez:2022mxx} for the equal mass prescription. With the equal $\widetilde m$ prescription of Eq. \eqref{wrong}, the double copy is obtained with the following representatives and coordinate space solutions
\begin{equation}
    \begin{array}{c@{\hspace{1cm}}c}
        \begin{aligned}
            f_{s=0, \tilde{m}} &= \frac{\braket{\lambda B}^{\tilde{m}}}{\braket{\lambda A} [\mu C]^{\tilde{m}+1}} \ , \\
            f_{s=1, \tilde{m}} &= \frac{\braket{\lambda B}^{\tilde{m}-1}}{\braket{\lambda A} [\mu C]^{\tilde{m}+2}} \ , \\
            f_{s=2, \tilde{m}} &= \frac{\braket{\lambda B}^{\tilde{m}-2}}{\braket{\lambda A} [\mu C]^{\tilde{m}+3}} \ ,
        \end{aligned}
        &
        \begin{aligned}
            \phi_0(x^{\alpha \dot{\alpha}}) &= - \frac{\braket{AB}^{\tilde{m}}}{(2v \cdot x)^{\tilde{m}+1}} \ , \\
            f_{\alpha\beta}(x^{\alpha \dot{\alpha}}) &= - A_\alpha A_\beta \frac{\braket{AB}^{\tilde{m}-1}}{(2v \cdot x)^{\tilde{m}+2}} \ , \\
            C_{\alpha\beta\gamma\delta}(x^{\alpha \dot{\alpha}}) &= - A_\alpha A_\beta A_\gamma A_\delta \frac{\braket{AB}^{\tilde{m}-2}}{(2v \cdot x)^{\tilde{m}+3}} \ ,
        \end{aligned}
    \end{array}
\end{equation}
where $\widetilde m$ is the mass of all the fields. Meanwhile in the prescription of Eq.~\eqref{correct}, the twistor representatives and coordinate space solutions for $\widetilde m_G>0$ are now given by 
\begin{equation} 
    \begin{array}{c@{\hspace{1cm}}c}
        \begin{aligned}
            f_{s=0, \widetilde{m}_\phi} &= \frac{\braket{\lambda B}^{\widetilde{m}_\phi}}{\braket{\lambda A} [\mu C]^{\widetilde{m}_\phi+1}} \, , \\
            f_{s=1, \widetilde{m}_\phi+1} &= \frac{\braket{\lambda B}^{\widetilde{m}_\phi}}{\braket{\lambda A} [\mu C]^{\widetilde{m}_\phi+3}} \, , \\
            f_{s=2, \widetilde{m}_\phi+2} &= \frac{\braket{\lambda B}^{\widetilde{m}_\phi}}{\braket{\lambda A} [\mu C]^{\widetilde{m}_\phi+5}} \, .
        \end{aligned}
        &
        \begin{aligned}
            \phi_0(x^{\alpha \dot{\alpha}}) &= - \frac{\braket{AB}^{\widetilde{m}_\phi}}{(2v \cdot x)^{\widetilde{m}_\phi+1}} \, , \\
            f_{\alpha\beta}(x^{\alpha \dot{\alpha}}) &= - A_\alpha A_\beta \frac{\braket{AB}^{\widetilde{m}_\phi}}{(2v \cdot x)^{\widetilde{m}_\phi+3}} \, , \\
            C_{\alpha\beta\gamma\delta}(x^{\alpha \dot{\alpha}}) &= - A_\alpha A_\beta A_\gamma A_\delta \frac{\braket{AB}^{\widetilde{m}_\phi}}{(2v \cdot x)^{\widetilde{m}_\phi+5}} \, .
        \end{aligned}
    \end{array}
\end{equation}
where $\widetilde{m}_\phi$ is the mass of the scalar field, $\widetilde{m}_\phi+1$ the mass of the spin one field, and $\widetilde{m}_\phi+2$ the mass of the spin two field. For $\widetilde m_G<0$ we have
\begin{equation}  \label{eq:DCtypeN}
    \begin{array}{c@{\hspace{1cm}}c}
        \begin{aligned}
            f_{s=0, \widetilde{m}_\phi} &= \frac{\braket{\lambda B}^{\widetilde{m}_\phi}}{\braket{\lambda A} [\mu C]^{\widetilde{m}_\phi+1}} \, , \\
            f_{s=1, \widetilde{m}_\phi-1} &= \frac{\braket{\lambda B}^{\widetilde{m}_\phi-2}}{\braket{\lambda A} [\mu C]^{\widetilde{m}_\phi+1}} \, , \\
            f_{s=2, \widetilde{m}_\phi-2} &= \frac{\braket{\lambda B}^{\widetilde{m}_\phi-4}}{\braket{\lambda A} [\mu C]^{\widetilde{m}_\phi+1}} \, .
        \end{aligned}
        &
        \begin{aligned}
            \phi_0(x^{\alpha \dot{\alpha}}) &= - \frac{\braket{AB}^{\widetilde{m}_\phi}}{(2v \cdot x)^{\widetilde{m}_\phi+1}} \, , \\
            f_{\alpha\beta}(x^{\alpha \dot{\alpha}}) &= - A_\alpha A_\beta \frac{\braket{AB}^{\widetilde{m}_\phi-2}}{(2v \cdot x)^{\widetilde{m}_\phi+1}} \, , \\
            C_{\alpha\beta\gamma\delta}(x^{\alpha \dot{\alpha}}) &= - A_\alpha A_\beta A_\gamma A_\delta \frac{\braket{AB}^{\widetilde{m}_\phi-4}}{(2v \cdot x)^{\widetilde{m}_\phi+1}} \, .
        \end{aligned}
    \end{array}
\end{equation}
where now we have that $\widetilde{m}_\phi-1$ is the mass of the spin one field, and $\widetilde{m}_\phi-2$ is the mass of the spin two field. Note only the case $\widetilde m_G<0$ keeps the bulk-to-boundary propagators equal (up to constant factors) for all spins. If we had considered the case where $\lambda^\alpha$ and $\mu^{\dot\alpha}$ are exchanged in the representative, as explained below Eq.~\eqref{Deltawave}, the bulk-to-boundary propagators would have remained equal for $\widetilde m_G>0$. Given this, we argue that the correct choice of representatives is as follows: for $\widetilde m_G<0$ the pole enclosed by the contour is $\la\lambda A\ra$ while for $\widetilde m_G>0$ the pole enclosed by the contour is $[\mu_x D]$.

Using these representatives, we can also construct an exact position-space double copy for massive pp-waves propagating on an AdS background. This is an exact solution since one can write pp-waves in Kerr-Schild form, as shown in the previous section, such that the equations of motion are linearised. Then, applying a covariant derivative of the full metric is equivalent to simply applying the AdS covariant derivative. Using the Penrose transform, we want to reproduce the Cotton tensor in Eq.~\eqref{eq:CottonPPwave} corresponding to the pp-waves metric Eq.~\eqref{eq:ppwave}. To do so, we need to consider the embedding of null Poincare coordinates which is given by
\begin{equation}
    X^{\alpha \dot{\alpha}} = L \begin{pmatrix}
        1 && u\\
        v && y_0^2+uv
    \end{pmatrix} \ .
\end{equation}
Taking $\braket{AB}=1$, $\widetilde m_G<0$,  $A_1=C_{\dot 1}=0$ and $A_0=L, C_{\dot{0}}=1$ and plugging it in Eq.\eqref{eq:DCtypeN} we have explicitly that the hyperbolic space fields are
\begin{equation}\label{ppwaveDCcoord}
    \begin{aligned}
        \phi_0&\,=- y_0^{\widetilde{m}_\phi+1}, \\
        f_{\alpha\beta}&\,= - A_\alpha A_\beta  y_0^{\widetilde{m}_\phi+1}, \\
        C_{\alpha\beta\gamma\delta}&\,= - A_\alpha A_\beta A_\gamma A_\delta y_0^{\widetilde{m}_\phi+1} \ ,
    \end{aligned}
\end{equation}
where one should remember that  $\widetilde{m}_V=\widetilde{m}_\phi-1$ is the mass of the spin one field, and $\widetilde{m}_G=\widetilde{m}_\phi-2$ is the mass of the spin two field. The general $u$ dependence in $f(u)$ in Eq.~\eqref{eq:ppwave} can be obtained by considering a non-constant $t_0$ in Eq.~\eqref{penrosebracket}. The solutions in Eq.~\eqref{ppwaveDCcoord} give the Cotton double copy for pp-waves of topologically massive theories in an AdS background which was first found in coordinate space in \cite{CarrilloGonzalez:2022mxx}.
%%%%%%%%%%%%%%%%%%%%
%%%%%%%%%%%%%%%%%%%%
\section{Type D double Copy in topologically massive gravity}\label{sectiontyped}
In this section, we will work in Lorentzian signature $(+,-,-)$, such that the comparison with the literature of exact solutions in TMG is simpler, and we obtain a black hole solution instead of the instantons that would arise in Euclidean signature. In this signature, the TMG action is
\begin{equation}
    S_{\text{TMG}} = \frac{1}{\kappa^2}\int\text{d}^3x\sqrt{g}
    \Bigg[-R-2 \Lambda -\frac{1}{2m_G}\epsilon^{\rho\mu\nu}\Gamma^{\sigma}_{\;\rho\lambda}
    \Big( \partial_\mu \Gamma^{\lambda}_{\;\sigma\nu}+\frac{2}{3}\Gamma^{\lambda}_{\;\mu\alpha}\Gamma^{\alpha}_{\;\nu\sigma}
    \Big)
    \Bigg] \ ,
    \label{eq:TMGaction}
\end{equation}
with equations of motion 
\begin{equation}
    G_{\mu\nu}+\frac{1}{ m_G}C_{\mu\nu}=\Lambda g_{\mu\nu} \ .
    \label{eq:TMGEOM}
\end{equation}
This theory allows for black hole solutions around AdS backgrounds, including the BTZ black hole \cite{BTZ1992,BTZ1993}. Here, we are particularly interested in black holes where the Cotton tensor is non-vanishing, which excludes the BTZ ones. Instead, we will focus on squashed AdS black holes \cite{Anninos:2008fx}. These black holes, analogously to the BTZ case, arise from the quotient of a biaxially squashed AdS spacetime by a discrete set of isometries\footnote{See \cite{Bieliavsky:2024hus} for a recent exploration on the geometry of warped AdS spacetimes}. This means that their Cotton tensor, which is only a local measure of the curvature, is given by the squashed AdS Cotton tensor. Furthermore, it is known that these squashed AdS spacetimes (locally) describe all the type D solutions of TMG \cite{Chow:2009km}. Type D solutions give rise to a gravitational field that behaves like that of an isolated gravitating object.  The squashed AdS solutions can be further classified as timelike-squashed AdS$_3$ or spacelike-squashed AdS$_3$. Here, we will focus on the latter, but the timelike-squashing can be obtained analogously.

More precisely, we will focus on the spacelike-squashed AdS$_3$ metric in Poincaré coordinates, whose line element takes the form  \cite{Chow:2009km}\footnote{Recall that we are using signature $(+,-,-)$, while \cite{Chow:2009km} uses $(-,+,+)$.}
\begin{equation}
    \text{d}s^2=\frac{9}{m_G^2-27\Lambda}\Bigg[ \frac{\text{d}t^2-\text{d}x^2}{x^2}-\frac{4m_G^2}{m_G^2-27\Lambda}\Bigg( \text{d}z+\frac{\text{d}t}{x} \Bigg)^2 \Bigg] \ ,
    \label{eq:ssads}
\end{equation}
where $\Lambda$ is the cosmological constant. This is a spacelike fibration over AdS$_2$ where the fiber is the real line. The factor
\begin{equation}
    \lambda\equiv\frac{4m_G^2}{m_G^2-27\Lambda} \ ,
\end{equation}
is the squashing parameter. For $\lambda>1$ we have stretching, for $\lambda<1$ squashing, and when $\lambda=1$ or equivalently $\Lambda=-m_G^2 /9$ the metric becomes the standard AdS. This metric has a spacelike Killing vector that will be useful in the Kerr-Schild construction and is given by
\begin{equation}
    l^\mu = \frac{m_G^2-27\Lambda}{6m_G}(0,0,1)^\mu \ .
    \label{eq:slkilling}
\end{equation}
It is easy to check that the Killing condition $\nabla_\mu l_\nu-\nabla_\nu l_\mu = 0$ is satisfied, and that this is a unit spacelike vector, $l_\mu l^\mu=-1$. The associated 1-form is described by
\begin{equation}
    l_\mu \text{d}x^\mu = \frac{6m_G}{m_G^2-27\Lambda}\Bigg(\frac{1}{x}\text{d}t+\text{d}z\Bigg) \ .
    \label{eq:slkillingoneform}
\end{equation}
Additionally, the traceless Ricci tensor, and hence the Cotton tensor of this metric, can be written in a simplified form by using this Killing vector
\begin{equation} \label{eq:tracelesRicciBH}
    S^{\mu\nu}=-\frac{C^{\mu\nu}}{m_G}=-\left(\frac{m_G^2}{9}+\Lambda \right)\left(g^{\mu\nu}-3l^\mu l^\nu\right) \ .
\end{equation}
From this we can see that the Lie derivative of $C^{\mu\nu}$ along the spacelike Killing vector $l^\mu$ vanishes
\begin{equation} \label{eq:LieCotton}
 \mathcal{L}_l C^{\mu\nu}\propto \mathcal{L}_l g^{\mu\nu} -3 \mathcal{L}_l l^\mu l^\nu=0  \ .
\end{equation}
The Jordan normal form of the traceless Ricci tensor can be computed by constructing a triad to re-express the traceless Ricci tensor in an orthonormal basis of the tangent bundle, $S^\mu_{\;\nu}\rightarrow S^A_{\;B}=e_\mu^{\;A}e^\nu_{\;B}S^\mu_{\;\nu}$.  One can solve for the triad in the usual way, to find
\begin{equation}
    e_A^{\;\mu}= \frac{x}{3}\sqrt{m_G^2-27\Lambda}\left(
\begin{array}{ccc}
 1 & 0 & -\frac{1}{x} \\
 0 & 1 & 0 \\
 0 & 0 & \frac{1}{2
   x }\sqrt{1-\frac{27 \Lambda}{m_G^2}}\\
\end{array}
\right)_A^{\;\;\mu}\ ,
    \label{eq:typedtriad}
\end{equation}
and applying the triad (and the co-triad) one finds
\begin{equation}
    S^A_{\;B}= -\left(\frac{m_G^2}{9}+\Lambda \right)
\left(
\begin{array}{ccc}
 1  & 0 & 0 \\
 0 & 1  & 0 \\
 0 & 0 & -2  \\
\end{array}
\right)^A_{\;\;B}.
    \label{eq:JNFtracelessricci}
\end{equation}
 
\subsection{Kerr-Schild metric for type D solutions of TMG}
The existence of the Kerr-Schild prescription for a generic base metric is not guaranteed, however, we can use our intuition to understand what choice of base metric will allow for the expression in Eq.~\eqref{eq:KSmetric} to exist. For example, AdS$_3$ may be written in Kerr-Schild form with a flat background metric, such that the cosmological constant is thought of as an exact perturbation generating the gravitational field. Similarly, we expect that biaxially-squashed AdS$_3$ may also be written in the form of \eqref{eq:KSmetric} with the base metric being AdS and the exact perturbation being now the squashing of the AdS length in one direction.  

In order to write the metric of the squashed AdS in Eq.~\eqref{eq:ssads} in Kerr-Schild form we must identify the background metric $\bar{g}_{\mu\nu}$, the scalar field $\phi$, and the Kerr-Schild vector $k_\mu$. Since the Kerr-Schild vector must be null and geodetic with respect to both the full and the base metrics, we start by finding $k_\mu$. Let $k^\mu=(N_0(t,x,z),N_1(t,x,z),N_2(t,x,z))^\mu$ be an arbitrary vector with the $N_i$ functions of the coordinates of the full metric. The null condition implies
\begin{equation}
    g_{\mu\nu}k^\mu k^\nu = 0 = -9\frac{3(9 \Lambda+m_G^2)N_0^2+(m_G^2-27 \Lambda)N_1^2+8xm_G^2N_0N_2+4x^2m_G^2N_2^2}{x^2(m_G^2-27 \Lambda)^2},
    \label{eq:nullcondition}
\end{equation}
where we have left out the functional dependences of the $N_i$ for brevity. The Kerr-Schild vector must also be geodetic. For this to be the case, a necessary condition is \cite{Stephani:2003tm}
\begin{equation}
S_{\mu\nu}k^\mu k^\nu=0 \ ,
\end{equation}
which in turn requires
\begin{equation}
    k\cdot l = 0 \ ,
\end{equation}
as seen by using Eq.~\eqref{eq:tracelesRicciBH}. This imposes the additional constraints
\begin{equation}
    -\frac{6m_G(N_0+xN_2)}{x (m_G^2-27 \Lambda)}=0\quad\Rightarrow\quad N_2=-\frac{1}{x}N_0,
    \label{eq:slkdotks}
\end{equation}
so that the Kerr-Schild vector is now of the form $k^\mu=(N_0,N_1,-N_0/x)^T$. Re-assessing the null condition now enforces $N_1=\pm N_0$, so that we have functional dependence on $N_0$, only,
\begin{equation}
    k^\mu=\Bigg(N_0,\pm N_0,-\frac{N_0}{x}\Bigg)^T.
\end{equation}
Having this simplified form, we now require that the Kerr-Schild vector satisfies the geodetic condition for the full metric as in Eq.~\eqref{eq:ksvec1}. Evaluating the geodetic condition yields two copies of the following condition (the $z$-component is already vanishing):
\begin{equation}
    0=k^\mu\nabla_\mu k_\nu \Rightarrow 0 = 2N_0\pm \partial_z N_0 - x(\partial_x N_0 \pm \partial_t N_0).
    \label{eq:geodeticpde}
\end{equation}
This partial differential equation is solvable and one finds
\begin{equation}
    N_0(t,x,z) = x^2 f(x\mp t,z\pm\log(\pm x)),
    \label{eq:n0solution}
\end{equation}
where $f$ is a scalar function of the listed arguments. We choose the upper signs for simplicity, so that
\begin{equation}
    N_0(t,x,z)=x^2f(x-t,z+\log(x)).
    \label{eq:N0sol}
\end{equation}

Next, we construct the background metric as
\begin{equation}
    \bar{g}_{\mu\nu}=g_{\mu\nu}-\phi k_\mu k_\nu \ .
    \label{eq:bgmetric}
\end{equation}
As yet, we have made no attempt to solve for the form of $\phi(t,x,z)$. However, we will make a simplifying assumption about the scalar field and take $\phi(t,x,z)=\phi(x)$. This simplification is made from physical expectations of the double copy construction. If the metric (the double copy) admits Killing vectors, then its Lie derivative along the Killing vectors vanishes. We can expect that the single copy and the zeroth copy have a similar property, which will be explored in detail below. In the Kerr-Schild double copy prescription the scalar $\phi$ gives the zeroth copy \cite{Monteiro:2014cda}. Hence, we require that the Lie derivatives of $\phi(t,x,z)$ along the Killing vectors vanish. In particular, for the current squashed AdS case one can find a timelike Killing vector $t^\mu$ in the direction of $t$ \textit{only}, as well as a spacelike Killing vector in the direction of $z$ \textit{only} (this is the vector in Eq.~\eqref{eq:slkilling}). The restrictions on $\phi$ are then $l^\mu\partial_\mu\phi=t^\mu\partial_\mu\phi=0$. This leaves the only possibility of non-trivial dependence of $\phi$ on $x$. Using \eqref{eq:bgmetric}, the background metric is initially of the form
\begin{equation}
   \bar{g}_{\mu\nu}= \frac{1}{x^2(m_G^2-27\Lambda)^2}\left(
\begin{array}{ccc}
 -27 \left(3 f^2 x^2 \phi+m_G
   ^2+9\Lambda\right) & 81 f^2 x^2 \phi
    & -36 m_G ^2 x \\
 81 f^2 x^2 \phi  & -9 \left(9
   f^2 x^2 \phi +m_G ^2-27
  \Lambda\right) & 0 \\
 -36 m_G ^2 x & 0 & -36 m_G ^2 x^2
   \\
\end{array}
\right)_{\mu\nu}\ ,
    \label{eq:bgmetricmatrix}
\end{equation}
where we have left out the argument(s) of $\phi$ and $f$ for brevity. Critically, the background metric will also satisfy the TMG equations of motion \eqref{eq:TMGEOM} with the tensors replaced by their `barred' equivalents (\textit{e.g.} $R\rightarrow\bar{R}$ to indicate that it is the Ricci scalar associated with the background metric $\bar{g}_{\mu\nu}$).

We now want to test our intuition and check that the base metric can be given by AdS. We start by looking at the Ricci scalar and requiring that it corresponds to that of a maximally symmetric spacetime with cosmological constant $\bar{\Lambda}$ such that
\begin{equation}
    \bar{R}=-6\bar{\Lambda}\quad \Rightarrow\quad -6 \Lambda - 
  x^3 f^2 \left(2 \phi' + x \phi''\right) = -6 \bar{\Lambda},
    \label{eq:bgricci}
\end{equation}
where the prime notation on the $\phi$'s indicates derivatives with respect to $x$. We can solve this partial differential equation for $\phi$ and find
\begin{equation}
    \phi(x)= \beta+\int_{1}^{x}\Bigg[ \frac{1}{p^2}\Bigg(\alpha +6\int_{1}^{p} \frac{\bar{\Lambda}-\Lambda}{q^2f(q-t,z+\log(q))^2}\text{d}q \Bigg)\Bigg]\text{d}p,
    \label{eq:phisol}
\end{equation}
where $\alpha$ and $\beta$ are constants. Note that $\beta$ sets the value of $\phi$ as $x\rightarrow\infty$, that is, deep in the bulk. Since maximally symmetric spacetimes are conformally flat, they have a vanishing Cotton tensor and, due to the TMG equations of motion, this also implies that the traceless Ricci tensor is vanishing,
\begin{equation}
    \bar{S}_{\mu\nu} \propto \bar{C}_{\mu\nu}=0 \ .
    \label{eq:tracelessricci}
\end{equation}
Explicit evaluation of the traceless Ricci tensor's $zz$ component yields
\begin{equation}
    \bar{S}_{zz}\propto 2m_G^2+18\Lambda+3x^3f^2(2\phi'+x\phi'') = 0.
    \label{eq:tracelessricci2}
\end{equation}
Using this equation together with \eqref{eq:bgricci} implies
\begin{equation}
 \bar{\Lambda}=-\frac{m_G^2}{9}.
    \label{eq:tracelessriccicondition}
\end{equation}
Note that this result is not surprising, if we were to set $\Lambda=\bar{\Lambda}$ as given in Eq.~\eqref{eq:tracelessriccicondition} then the squashing parameter of the squashed AdS metric would vanish. In other words, the base metric is the {\it unsquashed} metric and the Kerr-Schild factor gives the squashing.

In a similar vein, for the Cotton tensor's $zz$ component one finds
\begin{equation}
\begin{split}
    \bar{C}_{zz}=&-3 x^3 \left(m_G ^2-27 \Lambda\right)
   f(u,v) f^{(0,1)}(u,v) \left(x
   \phi ''+2 \phi '\right)\\
   &+6 m_G ^2
   x^3 f(u,v)^2 \left(x \phi ''+2
   \phi '\right)+4 m_G ^2 \left(m_G
   ^2+9\Lambda \right)=0 \ ,
\end{split}
    \label{eq:cottoncondition}
\end{equation}
where we have defined $u(x,t)=x-t$ and $v(x,z)=z+\log(x)$ . Using \eqref{eq:bgricci} and \eqref{eq:tracelessriccicondition} in \eqref{eq:cottoncondition} yields
\begin{equation}
   \frac{162 \left(3
   \Lambda+\bar{\Lambda}\right)
   \left(\bar{\Lambda}
   -\Lambda \right) f
   ^{(0,1)}(u,v)}{f (u,v)}=0\quad\Rightarrow\quad f(u,v)=f(u)=f(x-t),
    \label{eq:cottoncondition2}
\end{equation}
that is, $f$ is independent of its second argument. The restrictions in \eqref{eq:tracelessriccicondition} and \eqref{eq:cottoncondition2} are sufficient to guarantee the vanishing of all the components of $\bar{S}_{\mu\nu}$. However, there are remaining non-vanishing terms in the Cotton tensor. Explicitly, there are two equations that must simultaneously vanish to guarantee the vanishing of the Cotton tensor
\begin{equation}
    \begin{split}
        & \frac{\Big(\partial_uf(u)\Big) \left(-9 x^2 f(u)^2 \phi (x)+2
   m_G ^2-54 \Lambda\right)}{f(u)}=0,\\
        & \frac{\Big(\partial_uf(u)\Big) \left(9 x^2 f(u)^2 \phi
   (x)+m_G ^2-27 \Lambda\right)}{x f(u)}=0.
        \label{eq:finalcottonconditions}
    \end{split}
\end{equation}
Using the explicit form of $\phi(x)$ from \eqref{eq:phisol} one can find that the non-constant solutions are incompatible, and so we are forced to require that $f=\text{constant}$. Since the Kerr-Schild vector is directly proportional to $f$ while the scalar depends on $f^{-2}$ (the $\alpha$ and $\beta$ terms can be appropriately rescaled by $f^{-2}$ since they are arbitrary constants), we see that this is just a constant rescaling of the single and zeroth copies thus in the following we set $f=1$ for simplicity.

In summary, we have found that the biaxially squashed AdS spacetimes can be written in Kerr-Schild form (Eq.~\eqref{eq:KSmetric})  with an AdS base metric that reads
\begin{equation}
    \bar{g}_{\mu\nu} = \frac{1}{x^2 \left(\bar{\Lambda }+3
   \Lambda \right)^2} \left(
\begin{array}{ccc}
 x   (\alpha -\beta  x) &
     x (\beta 
   x-\alpha )-3 (\Lambda-\bar{\Lambda })  &
   4 x\bar{\Lambda } \\
   x (\beta 
   x-\alpha )-3 (\Lambda-\bar{\Lambda })  &   x (\alpha -\beta  x)+6 \Lambda -2 \bar{\Lambda
   }  &
   0 \\
 4 x \bar{\Lambda } & 0 & 4 x^2
   \bar{\Lambda } \\
\end{array}
\right)_{\mu\nu} \ .
\label{eq:bgmetricfinal}
\end{equation}
The Kerr-Schild vector is
 \begin{equation}
    k^\mu = \Big(x^2,x^2,-x \Big)^\mu \ ,
    \label{eq:ksvecsimple}
\end{equation}
and the scalar is
\begin{equation}
    \phi(x) = \frac{3(\bar{\Lambda}-\Lambda)}{  x^2}-\frac{\alpha
   }{x}+\beta.
    \label{eq:phisolsimple}
\end{equation}
The simplest solution corresponds to setting $\alpha=\beta=0$ where \begin{equation}
    \bar{g}_{\mu\nu} = \frac{1}{x^2 \left(\bar{\Lambda }+3
   \Lambda \right)^2} \left(
\begin{array}{ccc}
0 &
     -3 (\Lambda-\bar{\Lambda })  &
   4 x\bar{\Lambda } \\
   -3 (\Lambda-\bar{\Lambda })  &   6 \Lambda -2 \bar{\Lambda
   }  &
   0 \\
 4 x \bar{\Lambda } & 0 & 4 x^2
   \bar{\Lambda } \\
\end{array}
\right)_{\mu\nu} \ ,
\label{eq:bgmetricfinalNoAB}
\end{equation}
and this will be the case analysed for twistor space. For now, we keep $\alpha=\beta\neq 0$ and analyse how they affect the single and zeroth copies. One can note that in the case $\bar{\Lambda}=\Lambda$ the full and base metric are AdS with the same cosmological constant and thus the Kerr-Schild factor 
\begin{equation}
k^\mu k^\nu\phi =     g_{\mu\nu}-\bar{g}_{\mu\nu} = \left(
\begin{array}{ccc}
 \frac{(\beta  x-\alpha )}{16 \Lambda ^2 x} &
   \frac{ (\alpha -\beta  x)}{16 \Lambda ^2 x}
   & 0 \\
 \frac{(\alpha -\beta  x)}{16 \Lambda ^2 x} &
   \frac{ (\beta  x-\alpha )}{16 \Lambda ^2 x}
   & 0 \\
 0 & 0 & 0 \\
\end{array}
\right) \ ,
    \label{eq:metricdifference}
\end{equation}
must be a diffeomorphism. One can see this by noting that, choosing
\begin{equation}
    \xi_\mu=\left(
\begin{array}{c}
 \frac{c_1 e^z+c_2
   e^{-z}}{x}-\frac{ (\alpha  t-\beta  x
   (t-x))}{32 \Lambda ^2 x} \\
 \frac{ c_3+ c_1 e^z- c_2
   e^{-z}}{x}+\frac{ (2 \alpha  t+(t-x) (2 \alpha -\beta 
   (t+x)))}{64 \Lambda ^2 x} \\
 \frac{c_4+c_3 t}{x}-\frac{ (t-x)
   (\beta  (t-x) (t+2 x)-6 \alpha  t)}{192 \Lambda
   ^2 x} \\
\end{array}
\right)_\mu,
    \label{eq:diffeovector}
\end{equation}
where the $c_i$ are arbitrary constants, one finds
\begin{equation}
    \nabla_\mu \xi_\nu+\nabla_\nu\xi_\mu = g_{\mu\nu}-\bar{g}_{\mu\nu},
    \label{eq:diffeoproof}
\end{equation}
and it is important to note that covariant derivatives above are with those of the \textit{full} metric. Hence, $\alpha$ and $\beta$ arise as parameters of the diffeomorphism between $g_{\mu\nu}$ and $\bar{g}_{\mu\nu}.$ Nevertheless, for $\bar{\Lambda}\neq\Lambda$ this is not a diffeomorphism; it gives additional freedom allowed in the Kerr-Schild construction that can have implications in the double copy procedure, as we will see below.
%%%%%%%%%%%%%%
\subsection{Kerr-Schild double copy}  \label{subsec:KStypeD}
We have found that the spacelike-squashed metric \eqref{eq:ssads} has a Kerr-Schild form with an AdS base metric given by Eq.~\eqref{eq:bgmetricfinal}. We will now find the single and zeroth copies using the Kerr-Schild double copy prescription. 

\paragraph{Single Copy}
Using Eq.~\eqref{eq:KSsinglecopy} we find that the single copy is given as follows 
\begin{equation}
    A_\mu = \phi k_\mu = \frac{  x (\alpha -\beta 
   x)+3 (\Lambda-\bar{\Lambda }) }{ x^2 \left(3
   \Lambda+\bar{\Lambda } \right)}\Big(1,-1,0  \Big)_\mu.
    \label{eq:ksansatz}
\end{equation}
We can now solve for the electromagnetic field tensor via
\begin{equation}
    F_{\mu\nu}=\bar{\nabla}_\mu A_\nu-\bar{\nabla}_\nu A_\mu = \frac{\alpha   x+6 (\Lambda-\bar{\Lambda })
   }{x^3 \left(3 \Lambda+\bar{\Lambda }
   \right)}\left(
\begin{array}{ccc}
 0 & 1 & 0 \\
 -1 & 0 & 0 \\
 0 & 0 & 0 \\
\end{array}
\right)_{\mu\nu} \ ,
    \label{eq:EMTensor}
\end{equation}
as well as its Hodge dual  
\begin{equation} \label{eq:typeDstarF}
    \star F^\mu = -\left(\frac{6 (\Lambda-\bar{\Lambda })}{ x}+\alpha\right)l^\mu.
\end{equation}
In this case, we see that the parameter $\beta$ is just a gauge freedom since it does not contribute to the field strength. Meanwhile, $\alpha$ determines the behaviour deep in the bulk.

Just as in the double copy, we have that the Lie derivative of the Cotton tensor along the warped Killing vector vanishes, we expect that the same property holds for the single copy for the Hodge dual of the field strength. It is straightforward to see from \eqref{eq:typeDstarF} that indeed
\begin{equation} \label{eq:LieCotton2}
 \mathcal{L}_l \star F^{\mu}=0  \ .
\end{equation}
Then, making the identification $\mu^2 = -\gamma\bar{\Lambda}$ for arbitrary $\gamma$, and using the equations of motion \eqref{eq:MCSEOM}, the single copy has a source of the form
\begin{equation}
    j^\mu = \left(
\begin{array}{c}
 -6 \left(\Lambda -\bar{\Lambda }\right)
   \left(\bar{\Lambda }+3 \Lambda \right) \\
 0 \\
 -\frac{\left(\bar{\Lambda }+3 \Lambda \right)
   \left(6 \left(\sqrt{\gamma }-4\right)
   \left(\Lambda -\bar{\Lambda }\right)+\alpha 
   \left(\sqrt{\gamma }-2\right)  x\right)}{2 
   x} \\
\end{array}
\right)^\mu.
\end{equation}
Now, one can think of the source of the biaxially squashed AdS as the squashing of the AdS length (cosmological constant) in the directions orthogonal to the Killing vector $l^{\mu}$. In analogy, we can require that the source in the single copy is orthogonal to the spacelike Killing vector, then 
\begin{equation}
    0=j^\mu l_\mu=\frac{3 \left(\sqrt{\gamma }-2\right)
   \left(\bar{\Lambda }+3 \Lambda \right)
   \sqrt{-\gamma  \bar{\Lambda }} \left(\alpha  x+6 \left(\Lambda-
   \bar{\Lambda }\right)
   \right)}{ x \left(\gamma  \bar{\Lambda }+27
   \Lambda \right)}
    \label{eq:kjorthogonality}
\end{equation}
must hold for arbitrary values of the other parameters. The only value for $\gamma$ which ensures this is $\gamma=4$; then the dimensionless mass is given by
\begin{equation}
   \widetilde m_V=\widetilde m_G -1=2\ ,
\end{equation}
that is, the mass of the gauge field single copy satisfies the double copy relation in Eq.~\eqref{correctMass}. With this choice, the source is timelike. This is what we expect for the source of the single copy of the squashed AdS space / warped black hole. The gravitational source is just the squashed curvature, no NUT charges, and no rotation. Hence, we expect that the electromagnetic source is purely an electric density. 

Next, we want to isolate the electric field and magnetic field by choosing a reference timelike observer given by the unit timelike vector $u^\mu$ and making use of \cite{Tsagas:2004kv}
\begin{equation}
\begin{split}
    E^\mu =& F^{\mu\nu}u_\nu,\\
    B =& u_\rho \bar{\epsilon}^{\rho\mu\nu}F_{\mu\nu} \ .
\end{split}
    \label{eq:EandBfield}
\end{equation}
Our choice of unit timelike vector is 
\begin{equation}
   u^\mu = \frac{x \left(3 \Lambda+ \bar{\Lambda }
   \right)}{2 \sqrt{-\bar{\Lambda }}}\left(
\begin{array}{c}
 1 +\frac{x(\alpha- \beta x)}{2\left(3 \Lambda+ \bar{\Lambda }
   \right)}\\
 \frac{x(\alpha- \beta x)}{2\left(3 \Lambda+ \bar{\Lambda }
   \right)} \\
 -\frac{1}{x}\left(1+\frac{x(\alpha- \beta x)}{2\left(3 \Lambda+ \bar{\Lambda }
   \right)}\right) \\
\end{array}
\right)^\mu.
\end{equation}
Using this unit timelike vector in \eqref{eq:EandBfield} one finds
\begin{equation}
    \begin{split}
        E^\mu =& \frac{(6 (\Lambda- \bar{\Lambda}) +\alpha 
   x) \left(6(\Lambda - \bar{\Lambda})+x(\alpha-\beta x)\right)}{4
   \sqrt{-\bar{\Lambda}}}\left(
\begin{array}{c}
 1 \\
 -\frac{8 \bar{\Lambda}-x(\alpha-\beta x)}{6(\Lambda - \bar{\Lambda})+x(\alpha-\beta x)} \\
 -\frac{1}{x} \\
\end{array}
\right)^\mu\\
        B =& 0.
    \end{split}
    \label{eq:ExplicitEfield}
\end{equation}
It is easy to check that the electric field is orthogonal to the spacelike Killing vector \eqref{eq:slkilling}, and, as expected, there is no magnetic field.

\paragraph{Zeroth Copy}
Using Eq.~\eqref{eq:phisolsimple} in  Eq.~\eqref{eq:singlecopyeq} we find that the source is given by
\begin{equation}
    \mathcal{J}(x)= \frac{3 \left(\bar{\Lambda }-\Lambda
   \right) \left(\widetilde m_{G}^2-23 \bar{\Lambda
   }\right)}{x^2}-\frac{\alpha  \left(-19 \bar{\Lambda
   }+12 \Lambda +\widetilde m_{\phi}^2\right)}{x}-\alpha ^2 +\beta  \left(6 
   \Lambda -5  \bar{\Lambda
   } +\widetilde m_{\phi}^2\right).
    \label{eq:zerothcopysource}
\end{equation}
Proceeding as in the single copy case, we choose the mass of the scalar according to the prescription in Eq.~\eqref{correctMass}: 
\begin{equation}
    \widetilde m_{\phi}=\widetilde m_G-2=1 \ .
\end{equation}
With this choice of $\widetilde m_{\phi}$ the source terms become
\begin{equation}
    \mathcal{J}(x)=\frac{72 \left(\Lambda -\bar{\Lambda
   }\right) \bar{\Lambda }}{x^2}+\frac{4 \alpha  \left(5
   \bar{\Lambda }-3 \Lambda \right)}{x}+6 \beta(\Lambda-\bar{\Lambda}) 
    +\alpha^2 
    \label{eq:zerothcopysourcetwistormotivation}
\end{equation}
and the effective mass of the scalar vanishes
\begin{equation}
(\widetilde{m}_{eff})^2 = \widetilde m_{\phi}^2-1=0 \ .
\end{equation}
An interesting observation can be made for both the single and zeroth copies: setting $\bar{\Lambda}=\Lambda$ results in the electric field \eqref{eq:EandBfield} vanishing if $\alpha=0$; similarly for the source of the bi-adjoint scalar field.

%%%%%%%%%%%%
\subsection{Twistor double copy}
To connect the Kerr-Schild double copy to a twistor space formulation, we must find representative functions that result in a type D Cotton spinor under the integral Penrose transform. 

\subsubsection{Type D cohomology representatives}
A type D Cotton spinor has two distinct principal spinors, each with multiplicity two, we should therefore look for representative functions $f$ that have a third-order pole. Such poles can arise from inverse factors of the form $\la \lambda A\ra$ and $[\mu \widetilde{A}]$, where $A_{\alpha}$ and $\widetilde{A}_{\dot\alpha}$ are constant spinors with $A_0\neq0$ and $x^{1\dot\alpha}\widetilde{A}_{\dot\alpha}\neq0$ (if these latter restrictions are not made then these factors will not produce a pole).

If the twistor function $f$ has a third order pole at $\lambda_\alpha\propto A_\alpha$ we may separate it into its `regular' and `pole' factors as, $f\left(\mu^{\dot\alpha}_{x},\lambda_{\alpha}\right)=\la\lambda A\ra^{-3}g\left(\mu^{\dot\alpha}_{x},\lambda_{\alpha}\right)$, where $g$ is regular at the pole. We may then compute the Penrose transform of this representative, leaving $g$ arbitrary
\begin{equation}\label{dtransf}
    \Psi_{\alpha\beta\gamma\delta}(x)=\cint \la\lambda\rd\lambda\ra\lambda_\alpha\lambda_\beta\lambda_\gamma\lambda_\delta\,\frac{g\left(\mu^{\dot\alpha}_{x},\lambda_{\alpha}\right)}{\la\lambda A\ra^3}\,,
\end{equation}
where the contour encloses the triple pole and no other singularities of the integrand. Assuming that the spinor $A_\alpha$ is not proportional to $\iota_\alpha =(0,1)_{\alpha}$ we can perform the integral in the chart where $\lambda_\alpha = (1,\xi)_\alpha$ with $\xi\in\C$, resulting in
\begin{equation}
    \Psi_{\alpha\beta\gamma\delta}(x)=-\frac{12}{A_0^5}\left(g\iota_{(\alpha}\iota_{\beta}+\frac23g'\iota_{(\alpha}A_{\beta}+\frac{1}{12}g''A_{(\alpha}A_{\beta}\right)A_{\gamma}A_{\delta)}\,,
\end{equation}
where $g$, $g'$ and $g''$ denote the zeroth, single and double derivatives of $g\left(\mu^{\dot\alpha}_{x}(\xi),\lambda_\alpha(\xi)\right)$ with respect to $\xi$, evaluated at $\xi=A_1/A_0$. We immediately see that $A_\alpha$ is a principal spinor of the solution with multiplicity at least equal to two, implying that the solution is type II, III, D, or N. If $g=0$ when evaluated at $\xi=A_1/A_0$, the solution is either type III or type N, so we discard these cases and take $g\neq0$. Then, the field is specifically type D if
\begin{equation}\label{typedcond}
    \left(\iota_{(\alpha}\iota_{\beta)}+\frac23\frac{g'}{g}\iota_{(\alpha}A_{\beta)}+\frac{1}{12}\frac{g''}{g}A_{(\alpha}A_{\beta)}\right) = S_{\alpha}S_{\beta}\,,\quad S_\alpha\neq A_\alpha\,.
\end{equation}
Since $A_\alpha$ is not proportional to $\iota_\alpha$, any spinor can be written as a linear combination of the two. It is then straightforward to see that a necessary condition for the term in brackets above to factorise as in Eq. \eqref{typedcond} is 
\begin{equation}\label{dcond}
    {g'}^2=\frac{3}{4}g''g\quad\text{and}\quad g\neq0\,,
\end{equation}
at $\xi=A_1/A_0$ which gives the Penrose transform \eqref{dtransf} as
\begin{equation}
    \Psi_{\alpha\beta\gamma\delta}(x)=-\frac{12}{A_0^5}\,g\, S_{(\alpha}S_{\beta}A_{\gamma}A_{\delta)}\quad\text{where}\quad S_{\alpha}=\left(\iota_\alpha +\frac13\frac{g'}{g}A_\alpha\right)\,.
\end{equation}
Note that we have not required the spinor $A_{\alpha}$ to be a constant in $x$. Indeed, if the triple pole enclosed by the contour is due to inverse powers of $[\mu \widetilde{A}]$ in the twistor function, it will lead to a pole at $\xi = \widetilde{A}(x)_{1}/\widetilde{A}(x)_{0}$.

The type D condition Eq. \eqref{dcond} is solved by
\begin{equation}\label{soln}
    g\left(\mu^{\dot\alpha}_{x}(\xi),\lambda_\alpha(\xi)\right)=\frac{c_1(x)}{(\xi-c_2(x))^3}\,,
\end{equation}
where $c_{1,2}(x)$ are arbitrary functions of $x$.
This form of $g$ in Eq. \eqref{soln} is given by the rational twistor functions (up to overall constants)
\begin{equation}
    f(\mu^{\dot\alpha},\lambda_{\alpha})=\begin{cases}
        [\mu A]^{-3}[\mu B]^{-3}\quad &\widetilde{m}_G=3\,,\\
        \la\lambda A\ra^{-3}[\mu B]^{-3}\quad &\widetilde{m}_G=0\,,\\
        \la\lambda A\ra^{-3}\la\lambda B\ra^{-3}\quad &\widetilde{m}_G=-3\,,
    \end{cases}
\end{equation}
where the corresponding values of $\widetilde m_G$ are given, and $A_{\alpha}$, $A_{\dot\alpha}$, $B_{\alpha}$ and $B_{\dot\alpha}$ are constant spinors. While we have not claimed that \textit{only} functions of this form can provide type D solutions via the Penrose transform, it seemed difficult to find an alternative representative that would satisfy this.  Also, note that while one could naively expect to be able to obtain these representatives from the dimensional reduction of 4d type D representatives, this does not seem to be the case as shown in Appendix~\ref{app:dimred}. Explicitly, the Penrose transforms of each case are found to be
\begin{equation}\label{doublecopy}
    \begin{aligned}
        \widetilde m_G&=3\,,\qquad &&\Psi_{\alpha\beta\gamma\delta}=6\frac{A(x)_{(\alpha}A(x)_{\beta}B(x)_{\gamma}B(x)_{\delta)}}{[B A]^5},\\
        \widetilde m_G&=0\,,\qquad &&\Psi_{\alpha\beta\gamma\delta}=-3\frac{A_{(\alpha}A_{\beta}B(x)_{\gamma}B(x)_{\delta)}}{(v \cdot x)^5},\\
        \widetilde m_G&=-3\,,\qquad &&\Psi_{\alpha\beta\gamma\delta}=6\frac{A_{(\alpha}A_{\beta}B_{\gamma}B_{\delta)}}{\la B A\ra^5},\\
    \end{aligned}
\end{equation}
where $v_{\alpha\dot{\alpha}}=A_{\alpha}B_{\dot{\alpha}}$ and where we used the spinor identity $x^{\alpha\dot{\alpha}}u_{\alpha\dot{\beta}}= 2 (x \cdot u)\delta^{\dot\alpha}_{\dot\beta}-x_{\alpha\dot\beta}u^{\alpha\dot\alpha}$ while taking $x^2=1$.
For the cases $\widetilde
m_G=\pm3$, the above spinor fields are the spinor equivalent of the Cotton tensor of a spacetime that obeys the TMG equations of motion. The case $\widetilde m_G=0$ does not correspond to a solution of TMG and will be considered later in Section \ref{massless}. Note that the restriction to these values of the masses in fact coincides with what was found from the Kerr-Schild construction above. The base metric in the Kerr-Schild construction satisfies the same constraint on the mass as can be seen in Eq.~\eqref{eq:tracelessriccicondition}. Since type D solutions are locally squashed AdS, this simply tells us that the exact {\it perturbation} around AdS that generates the non-zero Cotton curvature tensor corresponds to the squashing around the unsquashed metric that has $\widetilde m_G=\pm3$.

%%%%%%%%%%%%%%%%%%%%%%%%%%%%%%%%%
\subsubsection{Type D Cotton double copy}
In the following, we will focus on the cases $\widetilde m_G=\pm3$ which are solutions of TMG. To physically interpret the above solutions it is useful to convert each one back to its tensor form using the inverse connection symbols (`Pauli matrices') corresponding to the metric signature $(+,-,-)$ given in Appendix \ref{3dspinor}
\begin{equation}
     C_{AB} = \frac{1}{4}\hs_A^{\alpha\beta}\hs_B^{\gamma \delta}\Psi_{\alpha\beta\gamma\delta}\,.
\end{equation}

We shall give explicit expressions for the case $\widetilde m_G=-3$ but the same steps follow for the $\widetilde m_G=3$ case. One can find a Lorentz transformation (which preserves the $(+,-,-)$ signature metric) given by:
\begin{equation} \label{eq:Ltransf}
    \Lambda_{A'}^{\phantom{A}B}=\left(
\begin{array}{ccc}
 -\frac{A_{0}^2+A_{1}^2+B_{0}^2+B_{1}^2}{2 A_{1} B_{0}-2 A_{0} B_{1}} & -\frac{A_{0}^2-A_{1}^2+B_{0}^2-B_{1}^2}{2 A_{1} B_{0}-2 A_{0} B_{1}} & -\frac{A_{0} A_{1}+B_{0} B_{1}}{A_{0} B_{1}-A_{1} B_{0}} \\
 -\frac{A_{0}^2+A_{1}^2-B_{0}^2-B_{1}^2}{2 A_{1} B_{0}-2 A_{0} B_{1}} & -\frac{A_{0}^2-A_{1}^2-B_{0}^2+B_{1}^2}{2 A_{1} B_{0}-2 A_{0} B_{1}} & -\frac{A_{0} A_{1}-B_{0} B_{1}}{A_{0} B_{1}-A_{1} B_{0}} \\
 -\frac{A_{0} B_{0}+A_{1} B_{1}}{A_{0} B_{1}-A_{1} B_{0}} & -\frac{A_{0} B_{0}-A_{1} B_{1}}{A_{0} B_{1}-A_{1} B_{0}} & -\frac{A_{0} B_{1}+A_{1} B_{0}}{A_{1} B_{0}-A_{0} B_{1}} \\
\end{array}
\right)
\end{equation}
which greatly simplifies the Cotton tensor to 
\begin{equation}
    {C}_{A'B'} = \frac{1}{2\la AB\ra^3}\begin{pmatrix}
        1 & 0 & 0\\
        0 & -1 & 0\\
        0 & 0 & 2
    \end{pmatrix}\quad\implies\quad   {C}^{A'}_{\phantom{A}B'} = \frac{1}{2\la AB\ra^3}\begin{pmatrix}
        1 & 0 & 0\\
        0 & 1 & 0\\
        0 & 0 & -2
    \end{pmatrix}
\end{equation}
where the prime indices denote the new Lorentz frame. This result is consistent with the form of type D solutions \cite{Chow:2009km}.  We can further identify the above with the corresponding Cotton tensor of the squashed AdS which can straightforwardly be obtained from Eq.~\eqref{eq:JNFtracelessricci}. Note that thanks to the Kerr-Schild nature of the squashed AdS metric we have the convenient property that $e^\nu_{\;A}e_\mu^{\;B}C^A_{\;B}=\bar{e}^\nu_{\;A}\bar{e}_\mu^{\;B}C^A_{\;B}$, where $\bar{e}^\nu_{\;A}$ and ${e}^\nu_{\;A}$ are the vielbeins of the squashed and base metric respectively. Thus, the result obtained from the full metric can be identified with the one here obtained with the base metric. This identification requires 
\begin{equation} \label{eq:identif}
\la AB \ra = \left[6 (-\bar{\Lambda})^{1/2}(\Lambda-\bar{\Lambda})\right]^{-\frac13}\,.
\end{equation} 

According to the prescription given in Eq. \eqref{correct}, since $\widetilde m_G=\pm 3$ we should take the single and zeroth copy representatives to have $\widetilde m_V=\pm 2$ and $\widetilde m_\phi=\pm 1$ respectively. By inspection, the algebraic structure of the Cotton spinor implies that the single copy and double copy have the representatives
\begin{equation}\label{singlecopy}
    \begin{aligned}
        f_{s=1,\widetilde m_V=2}&=\frac{1}{[\mu A]^2[\mu B]^2}\quad&&\implies\quad \phi_{\alpha\delta}=2\frac{A(x)_{(\alpha}B(x)_{\delta)}}{[A B]^3},\\
        f_{s=1,\widetilde m_V=-2}&=\frac{1}{\la \lambda A\ra^2\la\lambda B\ra^2}\quad&&\implies\quad \phi_{\alpha\delta}=2\frac{A_{(\alpha}B_{\delta)}}{\la A B\ra^3}.\\
    \end{aligned}
\end{equation}
and
\begin{equation}\label{zerothcopy}
    \begin{aligned}
        f_{s=0,\widetilde m_\phi=1}&=\frac{1}{[\mu A][\mu B]}\quad&&\implies\quad \phi=\frac{1}{[BA]},\\
        f_{s=0,\widetilde m_\phi=-1}&=\frac{1}{\la \lambda A\ra\la\lambda B\ra}\quad&&\implies\quad \phi=\frac{1}{\la BA\ra}.\\
    \end{aligned}
\end{equation}
Comparing Eq. \eqref{doublecopy}, \eqref{singlecopy} and \eqref{zerothcopy}, there is a manifest double copy structure in both twistor and position space
\begin{equation}
    \Psi_{\alpha\beta\gamma\delta} = \frac{3}{2}\frac{\phi_{(\alpha\beta} \phi_{\gamma \delta)}}{\phi}.
\end{equation}
It is interesting that a factor of $3/2$ appears. We do not have an explicit interpretation of this factor, but as mentioned in Section \ref{sec:DC} we can expect a dimensionful factor to be present in this equality. In this case, the dimensionful factor can depend on three scales: $m_G$, $m_V$, and $m_\phi$. Note that the algebraic type of the double copy was automatically preserved, i.e. the spin 1 fields must have 2 distinct principal spinors as opposed to a 2-fold principal spinor as expected from previous examples of classical double copies.

We now proceed to give a physical interpretation of the double copy. Recall that the Cotton spinor obtained from the twistor representative belongs to a warped black hole solution of TMG. As seen before, the Lie derivative along the spacelike Killing vector $l^\mu$ of the corresponding Cotton tensor vanishes. As in the Kerr-Schild double copy, here we expect that the Lie derivative along $l^\mu$ of the dual field strength single copy and the scalar zeroth copy also vanish. We will check this in the orthonormal frame (again after applying the Lorentz transformation from Eq.~\eqref{eq:Ltransf}) where we have\footnote{Note that the factor of $i$ appears due to our conventions of the sigma matrices in Appendix~\ref{ap:conventions}.}
\begin{equation}
    \star F_{A'}=-\frac{\ri}{\braket{AB}^2}\begin{pmatrix}0\\0\\1\end{pmatrix} =-\ri\left[6 (-\bar{\Lambda})^{1/2}(\Lambda-\bar{\Lambda})\right]^{\frac23}\begin{pmatrix}0\\0\\1\end{pmatrix}\ ,
\end{equation}
which is proportional to the Killing co-vector $l_A$ and so satisfies $\mathcal{L}_l\star F=0$ as required. In this case, the dual field strength tensor is given by $\star F^\mu=\bar{e}^\mu_{\;A}\star F^A$. As in Section \ref{subsec:KStypeD}, we can choose a unit-timelike vector $u^\mu=\bar{e}^\mu_{\;1}$ as a reference frame to find the electric and magnetic fields, which again will lead to a vanishing magnetic field. Finally, the zeroth copy is explicitly
\begin{equation}
    \phi = -\left[6 (-\bar{\Lambda})^{1/2}(\Lambda-\bar{\Lambda})\right]^{\frac13}\,.
\end{equation}
Since this is a scalar field with vanishing effective mass, it trivially satisfies the equation of motion $\bar{\nabla}^2\phi=0$. We can easily see that when $\Lambda=\bar{\Lambda}$, the zeroth, single, and double copy vanish. This is the limit in which the squashed metric becomes unsquashed. 

On the other hand, by keeping $\widetilde m$ the same, several options must be checked separately. Let's take $\widetilde m_V =-3$ for definiteness. The single copy representative has the form:
\begin{equation}\label{minus3EMrep}
    f_{s=1,\widetilde m_V=-3}=\frac{[\mu C]}{\la \lambda A\ra^n\la \lambda B\ra^{5-n}}\ , \qquad n=1,2\,,
\end{equation}
where $n=3,4$ are equivalent to making the swap $A \xleftrightarrow{} B$. Requiring the Petrov type to be preserved, we must manually choose $n=2$ ($n=1$ gives a 2-fold principal spinor, and therefore a wave-like solution). Requiring the two single copies to be identical implies that the scalar takes the form of
\begin{equation}
    \phi = \frac{[\mu C]^2}{\la \lambda A\ra \la \lambda B\ra^3},
\end{equation}
which is asymmetric in $A$ and $B$.
Alternatively, requiring a symmetric in $A, B$ zeroth copy
\begin{equation}
    \phi = \frac{[\mu C]^2}{\la \lambda A\ra^2 \la \lambda B\ra^2} \ ,
\end{equation}
leads to two different single copies given by Eq. \eqref{minus3EMrep} with both $n=2$ and $n=3$. 

In position space, these single copies are
\begin{equation}
\begin{aligned}
        n=2&\qquad\phi_{\alpha\beta}&&=\frac{1}{\la AB\ra^4}A_{(\alpha}\left(2B_{\beta)}\la AC\ra+A_{\beta)}\la BC\ra\right)\\
        n=3&\qquad\phi_{\alpha\beta}&&=\frac{1}{\la AB\ra^4}B_{(\alpha}\left(2A_{\beta)}\la CB\ra+B_{\beta)}\la CA\ra\right).
\end{aligned}
\end{equation}
Hence, with the equal mass prescription, neither possibility for the twistor double copy translates to a position space double copy. Furthermore, neither of these single copies give a dual field strength proportional to $l_A$ and so do not satisfy $\mathcal{L}_l\star F^\mu=0$.

%%%%%%%%%%%%%%%%%%%%
\section{The double copy from twistor space beyond type N and D} \label{sec:beyondtypesND}
Here, we will analyse the double copy for representatives that give a gravitational solution that is not Petrov type N or D. Note that any twistor double copy that is not type II, and hence does not have a Kerr-Schild form, only holds at linear order. In flat space, it was argued that only the double copy of highly algebraically special spacetimes is expected to be local (given by a product instead of a convolution) in twistor, position, and momentum space \cite{Luna:2022dxo}. This can be seen by obtaining the classical curvature tensors from scattering amplitudes and realising that only in these special cases, the twistor representatives can be fixed entirely by the three-point amplitudes. One can expect that a similar conclusion can be obtained for curved backgrounds in which case the double copy considered in this section would also fail to be local in position space. More precisely, what we mean by failure here is that the position space gravitational field obtained from a scalar and two spin 1 fields through the (local) Cotton double copy is not equal to the spin 2 position space field obtained through the twistor double copy. Note that in flat space, for beyond Type D solutions, the double copy becomes in general a sum of Weyl double copies \cite{Chacon:2021wbr}. Similar classical double copy constructions involving sums of squares arise in the presence of sources \cite{Easson:2021asd, Easson:2022zoh, Armstrong-Williams:2024bog}. Whether the cases examined below can be constructed in a similar manner remains an open question.

We will look at the example of type III in detail here. From Appendix \ref{app:petrov}, one can see that this requires that the gravitational representative has a second-order pole.
The simplest possibility that gives non-zero position space fields and respects the twistor double copy is found by fixing the pole structure as

\begin{equation}
    \begin{aligned}
        f_{s=0, \widetilde{m}_\phi=2}&=\frac{\braket{\lambda A}}{[\mu B]^2[\mu C]} \quad&&\implies\quad \phi&&= \frac{\braket{C(x)A}}{\braket{C(x)B(x)}^2}\\
        f_{s=1, \widetilde{m}_V=3}&=\frac{\braket{\lambda A}}{[\mu B]^2[\mu C]^3} \quad&&\implies\quad f_{\alpha \beta}&&= \frac{B(x)_{(\alpha}(2 C_{\beta)}\braket{B(x)A} + B(x)_{\beta)} \braket{C(x)A})}{\braket{C(x)B(x)}^4}\\
        f_{s=2, \widetilde{m}_G=4}&=\frac{\braket{\lambda A}}{[\mu B]^2[\mu C]^5}
    \end{aligned}
\end{equation}
\begin{equation}
\nonumber
    \begin{aligned}
\implies C(x)_{\alpha \beta \gamma \delta} \! &&= \frac{B(x)_{(\alpha} B(x)_{\beta}B(x)_{\gamma}(4 C(x)_{\delta)}\braket{B(x)A} + B(x)_{\delta)} \braket{C(x)A})}{\braket{C(x)B(x)}^6}.
    \end{aligned}
\end{equation}

This respects the twistor double copy, but not the position space double copy. Since the single copy has two non-degenerate principal null directions, its Cotton double copy gives a type D, on the other hand, the Cotton spinor arising from the twistor double copy is type III. Thus, the position space Cotton double copy prescription fails. To accommodate for the possibility of a double copy that is local in both twistor and position space, one should require two distinct single copies, one of which has a degenerate null principal direction (i.e. the representative has a simple pole). This constrains the most natural ansatz to be
\begin{equation}
    \begin{aligned}
        f_{s=0, \widetilde{m}_\phi=2}&=\frac{\braket{\lambda A}}{[\mu B][\mu C]^2} \quad&&\implies\quad \phi&&= \frac{\braket{AB(x)}}{\braket{C(x)B(x)}^2}\\
        f_{s=1, \widetilde{m}_V=3}&=\frac{\braket{\lambda A}}{[\mu B]^2[\mu C]^3} \quad&&\implies\quad f_{\alpha \beta}&&= \frac{B(x)_{(\alpha}(2 C_{\beta)}\braket{B(x)A} + B(x)_{\beta)} \braket{C(x)A})}{\braket{C(x)B(x)}^4}\\
        f'_{s=1, \widetilde{m}_V=3}&=\frac{\braket{\lambda A}}{[\mu B][\mu C]^4} \quad&&\implies\quad f'_{\alpha \beta}&&= \frac{B(x)_{(\alpha}B(x)_{\beta)} \braket{AB(x)}}{\braket{C(x)B(x)}^{4}}\\
        f_{s=2, \widetilde{m}_G=4}&=\frac{\braket{\lambda A}}{[\mu B]^2[\mu C]^{5}}
    \end{aligned}
\end{equation}
\begin{equation}
\nonumber
\begin{aligned}
    \implies C(x)_{\alpha \beta \gamma \delta}= \frac{4 \braket{B(x)A} B(x)_{(\alpha} B(x)_{\beta}B(x)_{\gamma} C(x)_{\delta)}}{\braket{C(x)B(x)}^6}+\frac{\braket{C(x)A} B(x)_{(\alpha} B(x)_{\beta}B(x)_{\gamma} B(x)_{\delta)} }{\braket{C(x)B(x)}^6},
\end{aligned}
\end{equation}
while the Cotton tensor obtained from the position space double copy is
\begin{equation}
    \frac{f_{(\alpha \beta}f'_{\gamma \delta)}}{\phi} =\frac{2 \braket{B(x)A} B(x)_{(\alpha} B(x)_{\beta}B(x)_{\gamma} C(x)_{\delta)}}{\braket{C(x)B(x)}^6}+\frac{\braket{C(x)A} B(x)_{(\alpha} B(x)_{\beta}B(x)_{\gamma} B(x)_{\delta)} }{\braket{C(x)B(x)}^6}.
\end{equation}
Surprisingly, the type N contribution is reproduced exactly and the type III contribution is off by a numerical factor of 2.  We can further consider the general $\tilde m$ case
\begin{equation}
    \begin{aligned}
        f_{s=0, \widetilde{m}_\phi}&=\frac{\braket{\lambda A}^{\widetilde{m}_\phi -1}}{[\mu B][\mu C]^{\widetilde{m}_\phi}} \\
        f_{s=1, \widetilde{m}_\phi+1}&=\frac{\braket{\lambda A}^{\widetilde{m}_\phi -1}}{[\mu B]^2[\mu C]^{\widetilde{m}_\phi+1}}\\
        f'_{s=1, \widetilde{m}_\phi+1}&=\frac{\braket{\lambda A}^{\widetilde{m}_\phi -1}}{[\mu B][\mu C]^{\widetilde{m}_\phi+2}}\\
        f_{s=2, \widetilde{m}_\phi+2}&=\frac{\braket{\lambda A}^{\widetilde{m}_\phi -1}}{[\mu B]^2[\mu C]^{\widetilde{m}_\phi+3}},
    \end{aligned}
\end{equation}
 which gives
 \begin{equation}
     \begin{aligned}
        \phi&= \frac{\braket{AB(x)}^{\widetilde{m}_\phi -1}}{\braket{C(x)B(x)}^{\widetilde{m}_\phi}}\\
        f_{\alpha \beta}&= \frac{\braket{B(x)A}^{\widetilde{m}_\phi -2} }{\braket{C(x)B(x)}^{\widetilde{m}_\phi +2}}B(x)_{(\alpha}(2 C_{\beta)}\braket{B(x)A} + (\widetilde{m}_\phi -1) B(x)_{\beta)} \braket{C(x)A})\\
         f'_{\alpha \beta}&= \frac{ \braket{AB(x)}^{\widetilde{m}_\phi-1}}{\braket{C(x)B(x)}^{\widetilde{m}_\phi+2}}B(x)_{(\alpha}B(x)_{\beta)}\\
        C(x)_{\alpha \beta \gamma \delta}&= \frac{\braket{B(x)A}^{\widetilde{m}_\phi-2}}{\braket{B(x)C(x)}^{\widetilde{m}_\phi+4}}B(x)_{(\alpha} B(x)_{\beta}B(x)_{\gamma}(4 C(x)_{\delta)}\braket{B(x)A} +(\widetilde{m}_\phi-1)B(x)_{\delta)} \braket{C(x)A}).
     \end{aligned}
 \end{equation}
Again, we have
\begin{equation}
    \frac{f_{(\alpha \beta}f'_{\gamma \delta)}}{\phi} =\frac{\braket{B(x)A}^{\widetilde{m}_\phi-2}}{\braket{B(x)C(x)}^{\widetilde{m}_\phi+4}}B(x)_{(\alpha} B(x)_{\beta}B(x)_{\gamma}(2 C(x)_{\delta)}\braket{B(x)A} + (\widetilde{m}_\phi-1)B(x)_{\delta)}\braket{C(x)A})  \ .
\end{equation}
As in the previous case, the position space Cotton double copy fails because of a factor of $2$ in the type III contribution. Interestingly, this factor turns out to be independent of $\widetilde{m}_\phi$. As we already mentioned, it was not expected that our twistor double copy would lead to a position space double copy. It is therefore very interesting that it fails by \textit{only} a factor 2. Furthermore, when the mass of the scalar is one, that is, it has vanishing effective mass, the second term drops and the position space and twistor double copies are proportional to each other as in type D.

For type II and type I, one can construct perfectly acceptable twistor double copies. However, it becomes increasingly hard to see if the twistor double copy can be taken such that the structure of the double copy is preserved in position space. Let us give a final example for an ansatz type II twistor double copy. First, we can again fix the $\lambda$ dependence as we did up to now so that all representatives scale as $f \propto \braket{\lambda A}^{\widetilde{m}_\phi-1}$. Next, a type II gravitational field representative should have a third-order pole. Physically, we should then require the spin 1 fields to have each a second-order pole at the same location. The rest of the twistor representative structure should respect the correct spin and mass homogeneity. The simplest and most symmetric solution leads to the following twistor double copy
\begin{equation}
    \begin{aligned}
        f_{s=0, \widetilde{m}_\phi}&=\frac{\braket{\lambda A}^{\widetilde{m}_\phi -1}[\mu C]^{\widetilde{m}_\phi+2}}{[\mu B] ([\mu D][\mu E])^{\widetilde{m}_\phi+1}} \\
        f_{s=1, \widetilde{m}_\phi+1}&=\frac{\braket{\lambda A}^{\widetilde{m}_\phi -1}}{[\mu B]^2[\mu D]^{\widetilde{m}_\phi+1}}\\
        f'_{s=1, \widetilde{m}_\phi+1}&=\frac{\braket{\lambda A}^{\widetilde{m}_\phi -1}}{[\mu B]^2[\mu E]^{\widetilde{m}_\phi+1}}\\
        f_{s=2, \widetilde{m}_\phi+2}&=\frac{\braket{\lambda A}^{\widetilde{m}_\phi -1}}{[\mu B]^3[\mu C]^{\widetilde{m}_\phi+2}},
    \end{aligned}
\end{equation}
which in general does not lead to a position double copy.

\section{Massless double copy}\label{massless}
In this section, we analyse the special case of the double copy of massless gauge fields. The massless limit of topologically massive Maxwell theory is simply Maxwell theory. In this case, the propagating degree of freedom is a scalar and not a spin one field. Meanwhile, the massless limit of topologically massive gravity is Chern-Simons theory, which has no propagating degrees of freedom in the bulk. We are again going to analyse the two options corresponding to the equal mass prescription in Eq.~\eqref{wrong} and the new prescription of Eq.~\eqref{correct}. We have already encountered the latter in Section \ref{sec:DC}, this is the case of the double copy of conserved currents where a massless gauge field divided by a scalar with vanishing effective mass double copies to the chiral point of TMG. In this case, a type N double copy is given by
\begin{equation} 
    \begin{array}{c@{\hspace{1cm}}c}
        \begin{aligned}
            f_{s=0, \widetilde{m}_\phi=-1} &= \frac{1}{\braket{\lambda A} \braket{\lambda B} } \, , \\
            f_{s=1,\widetilde{m}_V=0} &= \frac{1}{\braket{\lambda A} \braket{\lambda B} [\mu C]^{2}} \, , \\
            f_{s=2, \widetilde{m}_G=1} &= \frac{1}{\braket{\lambda A} \braket{\lambda B} [\mu C]^{4}} \, .
        \end{aligned}
        &
        \begin{aligned}
            \phi_0(x^{\alpha \dot{\alpha}}) &= - \frac{1}{\braket{AB}} \, , \\
            f_{\alpha\beta}(x^{\alpha \dot{\alpha}}) &= - A_\alpha A_\beta \frac{1}{\braket{AB}(2v \cdot x)^{2}} \, , \\
            C_{\alpha\beta\gamma\delta}(x^{\alpha \dot{\alpha}}) &= - A_\alpha A_\beta A_\gamma A_\delta \frac{1}{\braket{AB}(2v \cdot x)^{4}} \, .
        \end{aligned}
    \end{array}
\end{equation}
since the scalar is constant, this gives wave solutions of chiral gravity as the literal square of Maxwell waves. 

The other option is to consider the equal mass prescription in Eq.~\eqref{wrong}. While this immediately seems like an odd choice due to the mismatch in degrees of freedom (one in the vector and zero in the graviton), we analyse if there is any possible physical interpretation.  Note that in the massless case, the twistor representatives formally resemble the usual 4d representatives of flat space in the sense that $\lambda$ and $\mu$ scale in the same way with a total homogeneity of $-2s-2$. Thus one can expect that the prescription of \cite{White:2020sfn} (i.e. where all the fields are massless) leads to a physical double copy.

First, we need to clarify the spinors that enter the double copy in this limit. Since the traceless Ricci tensor and the Cotton tensor are proportional to each other in topologically massive gravity, Eq. \eqref{tmgspinor} can always be interpreted as the equation obeyed by either tensor up to a constant dimensionful factor. Up until now, we focused on the Cotton tensor as the analogue of the Weyl tensor in 3d. Nevertheless, this tensor is zero in Chern-Simons gravity due to the equations of motion. We can instead interpret the spinor arising from the Penrose transform as the traceless Ricci tensor since it satisfies the RHS of Eq.~\eqref{nonzerospin} thanks to the Bianchi identity and Eq.~\eqref{eq:cottonDef} when the Cotton tensor vanishes.

With this choice we have a double copy that squares the dual field strength of Maxwell to the traceless Ricci of a conformally flat metric, that is a solution of Chern-Simons gravity. Since the degrees of freedom do not match then we expect some type of topological information to be double copied. 

Assuming that the bulk degrees of freedom in the Maxwell theory are not excited, the field strength can be of purely topological nature. In fact, in 3d Maxwell theory, there is a topological $U(1)$ current given by \cite{tong}
\begin{equation}
    J^{\mu}_{Maxwell} = \frac{1}{4 \pi} \epsilon^{\mu\nu\rho}F_{\nu \rho},
\end{equation}
which is conserved thanks to the Bianchi identity and coincides with our single copy (up to a constant factor). This $U(1)$ topological symmetry acts on monopole operators.  Now, on the gravitational side, we can also construct a conserved current related to the double copy. Since the traceless Ricci is divergenceless, the current given by
\begin{equation}
    J^{\mu}_{\text{CS}} = S^{\mu \nu}l_{\nu} \ ,
\end{equation}
where $l^{\mu}$ is a Killing vector, is covariantly conserved \cite{Kastor:2004jk}.  Since Chern-Simons gravity is topological, the origin of this current is also purely topological. Thus the massless twistor double copy gives, in this equal mass case, a double copy in coordinate space between topological currents, up to the contraction with a Killing vector. Note that while the gauge and gravitational theories can have a topological interpretation, this will not necessarily be the case for the scalar. In principle, one could hope that the scalar is given by a topological soliton configuration, but while $m_\phi=0$, the effective mass is non-zero on the fixed AdS background due to the non-minimal coupling which leads to a trivial vacuum manifold given by $\phi=0$. Perhaps considering solutions to the full bi-adjoint scalar theory, instead of the linearized version can lead to a topological soliton solution, but these are also generically forbidden due to Derrick's theorem \cite{Derrick:1964ww}. 

%%%%%%%%%%%%%%%%%%%%
\section{Conclusions}

Since the Kerr-Schild double copy was discovered, there have been many developments towards understanding how a classical version of the double copy is realised beyond the well-known perturbative scattering amplitudes double copy in momentum space. One of the biggest open questions is whether a double copy relation holds in curved backgrounds. Here, we have tackled this question around fixed hyperbolic/AdS backgrounds from the perspective of twistor space.

We constructed a double copy for \cech cohomology class representatives in twistor space and provided a prescription for how the masses, or equivalently the conformal dimensions, of different spinning fields appearing in the double copy are related.  These cohomology class representatives give rise, through the Penrose transform, to solutions of the equations of motion of free massive particles propagating one helicity mode in AdS and in the spinless case to a non-minimally coupled massive scalar. The double copy gives a spin $2s$ solution as the square of a spin $s$ divided by a scalar. Additionally, in the spin one and spin two case, these equations also describe the equations of motion for the curvature tensors of topologically massive Yang-Mills and gravity.  Hence, we constructed a double copy not only for topologically massive classical solutions, but also for bulk-to-boundary propagators. Besides the double copy in twistor space and its corresponding Cotton double copy in position space, we also constructed the Kerr-Schild double copy in position space for type N and type D solutions.

We have shown that our double copy prescription has several interesting features. One of them allows us to keep the denominator of bulk-to-boundary propagators constant through the double copy prescription which can be seen as an analogue of the denominator of propagators being kept constant in the usual BCJ double copy for scattering amplitudes. While we only showed the explicit case of the bulk-to-boundary propagator, in principle one could obtain higher point correlators where one operator is in the bulk and the rest in the boundary from the Penrose transform. Whether these will also satisfy a local double copy in position space is unclear and will be considered in future explorations of a double copy for correlation functions in AdS$_3$. In the context of AdS/CFT, our double copy can also be seen to give rise to a double copy relation between the dual currents as we approach the conformal boundary. Furthermore, it takes conserved currents to conserved currents, which in this case are dual to a bulk massless gauge field and in the gravitational case to the chiral point of TMG.
 
Regarding the twistor double copy of topologically massive theories around hyperbolic/AdS space, we focused on type N and type D fields because these represent exact solutions but the double copy remains true at linear order for the other Petrov types. For instance, we gave an explicit construction of a type III double copy that becomes local in position space when considering scalars whose effective mass vanishes.

When looking at exact solutions, we found a local Cotton double copy in position space for both type N and type D as well as a Kerr-Schild version of the double copy. For type N, the Kerr-Schild double copy requires the mass prescription in Eq.~\eqref{correctMass} to obtain physical solutions of waves with vanishing sources. Meanwhile, the Cotton double copy can be constructed with both the novel Eq.~\eqref{correctMass} prescription as well as the equal mass prescription. For the type D case, first we constructed the Kerr-Schild form of the squashed AdS metrics, which represent all the type D solutions of TMG. To the best of our knowledge, this construction is not found elsewhere in the literature for an AdS base metric. This construction showed that the AdS length of the base metric has to be related to the graviton mass as $m_G \bar{L}=\pm3$. Similarly, from twistor space, we found that writing type D representatives required the same restrictions on the graviton mass. Note that this does not mean that the value of the mass is fixed, but that the value of the AdS length of the base metric is fixed. In other words, $m_G L$, is not fixed, where $L=-1/\Lambda$ gives the cosmological constant of the squashed AdS metric. 

In both the Kerr-Schild and the Cotton double copy we find that the double copy prescription of Eq.~\eqref{correct} gives what we consider a {\it physical} result, contrary to the equal mass prescription. When we perform the double copy using Eq.~\eqref{correct}, we find in both classical double copies that the single copy has a dual field strength whose Lie derivative along the spacelike Killing vector of the squashed AdS metric vanishes. This is in analogy to the Lie derivative along the spacelike Killing vector of the Cotton tensor vanishing, thus the symmetries are preserved. This breaks down when considering the equal mass prescription. Note that the type D double copy can also be interpreted as a double copy for warped black hole solutions (which are locally squashed AdS). In previous classical double copy constructions, the single copy of black holes gives the field strength of the corresponding charged black hole. It would be interesting to understand if this is also the case for our solution, that is, if the single copy corresponds to the dual field strength of the charged warped black hole solutions similar to those in \cite{Corral:2024xfv,Dereli:2000hz}.

We also note that the type D twistor double copy translates to a position space double copy up to a factor of $3/2$. A similar discrepancy between the twistorial double copy and its position space version was observed in type III. No rescaling of the representatives can be done to fix this factor. We highlighted that the Cotton double copy is missing a dimensionful factor in the relation, and since we have different scales in play, this could in principle explain the numerical factors seen in these examples. Although, it is interesting to note that this discrepancy is extremely constrained and is independent of the mass in the type III case. This suggests it is also independent of the masses in the general case.

Another interesting feature of the 3d TMG double copy arises when considering the usual starting point of the gauge theory. If we start with a gauge theory solution whose field strength has a vanishing Lie derivative along a given direction $l^\mu$. We can construct its Cotton double copy to obtain the gravitational solution, but we can go further and obtain the full metric since we have knowledge of a Killing vector of the solution \cite{Gurses:2010sm}.

Finally, the massless limit of the double copy constructed here poses an apparent paradox since the degrees of freedom between the single copy and the gravitation fields do not match. To solve this issue, we argued that, in this limit, the double copy is constrained to the topological sectors of the theories. Alternatively, we can consider again the prescription in Eq.~\eqref{correct} in which case the double copy is the chiral point of TMG. 

The results presented here open many interesting questions related to the double copy in curved spacetimes. While there are many attempts at constructing a double copy for correlators in (A)dS backgrounds, most suffer from issues at 4-points, although counterexamples, with well-defined double copies, exist \cite{Zhou:2021gnu, Alday:2021odx,Alday:2022lkk}. It would be interesting to analyse if this is also the case in topologically massive theories or if simplifications of the three-dimensional theories can give a physical double copy. It would also be compelling to explore in more detail the realisation of the double copy relation on the dual theory at the boundary. Last, the simplicity of topologically massive theories in AdS might allow us to recover the kinematic algebra in a similar manner to the one in self-dual theories in curved spacetimes \cite{CarrilloGonzalez:2024sto,Lipstein:2023pih}. We will explore the latter in future work.

%%%%%%%%
\section*{Acknowledgements}
We would like to thank Tim Adamo for discussions related to the minitwistor of Lorentzian AdS$_3$ and Sean Seet and David Skinner for discussions on minitwistor space and dimensional reductions. MCG is supported by the Imperial College Research Fellowship. TK is supported by an STFC studentship. CB is supported by the Oppenheimer Memorial Trust Research Fellowship and the Harry Crossley Research Fellowship. The research of SJ is supported by the ERC (NOTIMEFORCOSMO, 101126304) funded by the European Union, as well as the Simons Investigator award 690508. Views and opinions expressed are however those of the author(s) only and do not necessarily reflect those of the European Union or the European Research Council Executive Agency. Neither the European Union nor the granting authority can be held responsible for them.
\newpage
\appendix
\section{Conventions}
\label{ap:conventions}
%%%%%%%%%%%%%%%%%%%%%%%
\subsection{Spinors in 4 dimensions}
We will use the spacetime signature $(+,-,-,-)$ for real coordinates. So, $x\cdot x =t^2-x^2-y^2-z^2$ on flat real Minkowski. We will also use complex coordinates, in which case $x\cdot x$ is still given by the above, even though there is no longer a notion of signature. Complex coordinates are dimensionless in both three and four dimensions, whereas real four dimensional coordinates are given a length scale through the redefinition $x\xr x/L$ with $L\in\mathbb R^+$.

Due to the isomorphism of the complexified Minkowski isometry algebra $\mathfrak{so}(4,\mathbb{C})\cong \mathfrak{sl}(2,\C)\times\mathfrak{sl}(2,\C)$, Lorentz vector indices may be replaced by pairs of spinor indices corresponding to distinct SL$(2,\C)$ representations. The spinor equivalent of Lorentz tensors are defined by contracting with Pauli matrices as follows
\begin{equation}
    T^{\alpha\dot\alpha\ldots}_{\phantom{\alpha\dot\alpha\ldots}\beta\dot\beta\ldots}=\sigma^{\alpha\dot\alpha}_{A}\ldots\sigma_{\beta\dot\beta}^{B}T^{A\ldots}_{\phantom{A\ldots}B\ldots}\,,
\end{equation}
where $A, B$ are flat Lorentz indices and where the Pauli matrices are taken to be
\begin{equation}
    \sigma_A^{\alpha\dot\alpha}=\left\{\begin{pmatrix}
        1 & 0\\
        0 & 1
    \end{pmatrix},\begin{pmatrix}
        0 & 1\\
        1 & 0
    \end{pmatrix},\begin{pmatrix}
        0 & -\ri \\
        \ri & 0
    \end{pmatrix},\begin{pmatrix}
        1 & 0\\
        0& -1
    \end{pmatrix}\right\}\,,
    \label{sigma4d}
\end{equation}
with the conjugates. Spinor indices are raised and lowered with the epsilon symbol as follows
\begin{equation}
    \begin{aligned}
        a_\alpha=a^{\beta}\epsilon_{\beta\alpha}\,,\quad b^{\alpha}=\epsilon^{\alpha\beta}b_{\beta}\, \\
        \epsilon_{01}=1 = - \epsilon_{10}\,,\quad\epsilon^{01}=1=-\epsilon^{10}\,,
    \end{aligned}
\end{equation}
and similarly for the dotted version of these equations. Inner products between spinors are defined as
\begin{equation}
\begin{aligned}
        \la\lambda B\ra &=\lambda^{\alpha}B_{\alpha}= \lambda_1 B_0-\lambda_0 B_1\\
        \left[\mu A\right] &=\mu^{\dot\alpha} A_{\dot\alpha}= \mu_{\dot 1}A_{\dot 0}- \mu_{\dot 0}A_{\dot 1}=\mu^{\dot 1}A^{\dot 0}- \mu^{\dot 0}A^{\dot 1}.
\end{aligned}
\end{equation}

The dual matrices are defined by raising and lowering the vector and spinor indices by applying the Minkowski metric and epsilon symbol as appropriate. The spacetime coordinate spinor is then
\begin{equation}\label{minkcoord}
    x^{\alpha\dot\alpha}=\sigma_\mu^{\alpha\dot\alpha}x^\mu=\begin{pmatrix}
        x^0+x^3 & x^1-\ri x^2\\
        x^1+\ri x^2 & x^0-x^3
    \end{pmatrix}\,,
\end{equation}
 and the determinant gives the usual metric inner product
\begin{equation}
    \text{det}(x^{\alpha\dot\alpha})=t^2-x^2-y^2-z^2\,.
\end{equation}
To convert back to Lorentz indices just contract with the dual matrices with a factor of half per Pauli matrix, e.g.
\begin{equation}
    x^\mu=\frac12\sigma_{\alpha\dot\alpha}^{\mu}x^{\alpha\dot\alpha}\,.
\end{equation}
In general, if contracting between $n$ pairs of spinor indices, there will be a factor of $2^n$ relative to the contraction of $n$ spacetime indices.
For instance
\begin{equation}
\begin{aligned}
x^{\alpha\dot\alpha}x_{\alpha\dot\alpha}=\sigma_{\alpha\dot\alpha}^\mu x_\mu \sigma^{\alpha\dot\alpha}_\nu x^\nu =2 x^2\,,\\
    x^\mu x_\mu = \frac14 \sigma^\mu_{\alpha\dot\alpha}x^{\alpha\dot\alpha}\sigma_\mu^{\beta\dot\beta}x_{\beta\dot\beta}=\frac12 x^{\alpha\dot\alpha}x_{\alpha\dot\alpha},
\end{aligned}
\end{equation}
using that $2\delta_\alpha^{\beta}\delta_{\dot\alpha}^{\dot\beta}=\sigma^\mu_{\alpha\dot\alpha}\sigma_\mu^{\beta\dot\beta}$.

%%%%%%%%%%%%%%%%%%%

\subsection{Spinors in 3 dimensions}\label{3dspinor}
We denote 3d vielbein/Lorentz indices by capital letters at the beginning of the Latin alphabet. The spinor indices will be denoted by the beginning of the Greek alphabet ($\alpha, \beta , \dots$) and the spacetime indices will be denoted by the middle of the Greek alphabet ($\mu, \nu , \dots$). Many useful results and further explanations regarding spinors in three-dimensions may be found in \cite{castillo}.

A 2-1 homomorphism exists between the complexified 3d Lorentz group SO$(3,\mathbb C)$ and SL$(2,\mathbb C)$ and so vector indices can be replaced by pairs of symmetric spinor indices $\alpha, \beta$ (both un-dotted) using the Pauli/connection matrices
\begin{equation}
    \sigma_A^{\alpha\beta}=\sigma_A^{(\alpha\beta)}=\left\{\begin{pmatrix}
        1 & 0\\
        0 & 1
    \end{pmatrix},\begin{pmatrix}
        \ri & 0\\
        0& -\ri
    \end{pmatrix},\begin{pmatrix}
        0 & -\ri\\
        -\ri & 0
    \end{pmatrix}\right\}\,.
    \label{pauli3d}
\end{equation}
These satisfy the Clifford algebra
\begin{equation}
    (\sigma_{A})^{\alpha}_{\beta}(\sigma_{B})^{\beta}_{\gamma}+(\sigma_{B})^{\alpha}_{\beta}(\sigma_{A})^{\beta}_{\gamma}=2\eta_{AB}\delta^{\alpha}_{\gamma}\,,
\end{equation}
where $\eta_{AB}=\text{diag}(-1,-1,-1)$. In the main text we shall also use Pauli matrices that satisfy the Clifford algebra with $\eta_{AB}=\text{diag}(1,-1,-1)$. We denote these matrices with a hat, $\hs_A^{\alpha\beta}$. This gives
\begin{equation}
    \hs_A^{\alpha\beta}=\left\{\begin{pmatrix}
        \ri & 0\\
        0 & \ri
    \end{pmatrix},\begin{pmatrix}
        -\ri & 0\\
        0& \ri
    \end{pmatrix},\begin{pmatrix}
        0 & \ri\\
        \ri & 0
    \end{pmatrix}\right\}\,,
    \label{pauli3dp}
\end{equation}
and the duals
\begin{equation}
        \hs^A_{\alpha\beta}=\left\{\begin{pmatrix}
        \ri & 0\\
        0 & \ri
    \end{pmatrix},\begin{pmatrix}
        -\ri & 0\\
        0& \ri
    \end{pmatrix},\begin{pmatrix}
        0 & \ri\\
        \ri & 0
    \end{pmatrix}\right\}\,.
    \label{dpauli3dp}
\end{equation}

The relations between spinors and tensors are then
\begin{equation}
    T^{\alpha\beta}=\sigma^{\alpha \beta}_A T^{A}\,,\quad T_{\alpha\beta}=\sigma_{\alpha \beta}^A T_{A}\,,\quad \sigma_{\alpha\beta}^A\sigma_{B}^{\alpha\beta}=-2\delta^{A}_{B},
    \label{vectospin}
\end{equation}
\begin{equation}
    T_A = \frac12\sigma^{\alpha\beta}_A T_{\alpha\beta}\,,\quad T^A = \frac{1}{2}\sigma_{\alpha\beta}^{A}T^{\alpha\beta}\,,\quad T_{\alpha\beta}U^{\alpha\beta}=-2T_AU^A=-2T\cdot U\,.
    \label{spinvector3d}
\end{equation}
A few useful identities are (with the three dimensional Levi-Civita symbol satisfying $\epsilon_{012}=1$)
\begin{equation}
    \epsilon_{\alpha\beta}\epsilon_{\gamma\delta}+\epsilon_{\beta\gamma}\epsilon_{\alpha\delta}+\epsilon_{\gamma\alpha}\epsilon_{\beta\delta}=0\,,
\end{equation}
\begin{equation}
    \eta_{\alpha\beta\gamma\delta}=-(\epsilon_{\alpha\gamma} \epsilon_{\beta\delta} + \epsilon_{\alpha\delta}\epsilon_{\beta\gamma})\,,
\end{equation}
\begin{equation}
   \epsilon_{\alpha\beta\gamma\delta\zeta\eta}=-2(\epsilon_{\alpha\gamma}\epsilon_{\beta\zeta}\epsilon_{\delta\eta}+\epsilon_{\beta\delta}\epsilon_{\alpha\eta}\epsilon_{\gamma\zeta})\,.
\end{equation}

%%%%%%%%%%%%%%%%%%%
\section{Scalar equation of motion from the Penrose transform} \label{ap:scalar}
In this Appendix, we show one direction of the equivalence in Eq. \eqref{zerospin}. We will start from the integral representation of the Penrose transform and verify the equation of motion satisfied by the scalar. First, using Eq. \eqref{covariantderivative}, we note that when all indices are contracted, the connection terms vanish such that
\begin{equation}
    \nabla_{\h\ }^{\alpha \beta} \psi_{\alpha \beta} = \epsilon^{\alpha \gamma} \epsilon^{\beta \delta} 2x^{\Dot{\beta}}_{(\gamma}\partial_{\delta)\Dot{\beta}} \psi_{\alpha \beta} \ .
\end{equation}
We can now therefore easily apply the Laplacian on $f\equiv f_{s=0,\widetilde{m}_\phi}(\mu,\lambda)$ to obtain
\begin{equation}
    \nabla_{\h\ }^{\alpha \beta}\nabla_{\h\, \alpha \beta} f = -2 x^{\alpha \Dot{\gamma}}\partial_{\beta \Dot{\gamma}} (x^{\Dot{\beta}}_{\alpha
    }\partial^{\beta}_{\Dot{\beta}}f) -2 x^{\alpha \Dot{\gamma}}\partial_{\beta \Dot{\gamma}} (x^{\beta \Dot{\beta}}\partial_{\alpha \Dot{\beta}}f) \ .
\end{equation}
Using the homogeneity of $f$, we obtain the following
\begin{equation}
\begin{aligned}
    \mu^{\Dot{\alpha}} \frac{\partial f}{\partial \mu^{\Dot{\alpha}}} = -(1+\widetilde{m}_\phi)f 
    \implies \mu^{\Dot{\alpha}} \mu^{\Dot{\beta}} \frac{\partial^2 f }{\partial \mu^{\Dot{\alpha}} \partial \mu^{\Dot{\beta}}} = (1 + \widetilde{m}_\phi)(2+ \widetilde{m}_\phi) f \ .
\end{aligned}
\end{equation}
Then using the incidence relation
\begin{equation}
    \frac{\partial}{\partial x^{\alpha \Dot{\beta}}} = \frac{\partial \mu^{\Dot{\gamma}}}{\partial x^{\alpha \Dot{\beta}}} \frac{\partial}{\partial \mu^{\Dot{\gamma}}} = \lambda_{\alpha} \frac{\partial}{\partial \mu^{\Dot{\beta}}}\ ,
\end{equation}
we obtain the scalar equation of motion 
\begin{equation}
\begin{aligned}
   \nabla_{\h\ }^{\alpha \beta}\nabla_{\h\, \alpha \beta} \phi = \oint \langle \lambda d \lambda \rangle  \nabla_{\h\ }^{\alpha \beta}\nabla_{\h\, \alpha \beta} f 
   &\,= \oint \langle \lambda d \lambda \rangle ( -6 \mu^{\Dot{\beta}} \frac{\partial f}{\partial \mu^{\Dot{\beta}}} - 2 \mu^{\Dot{\alpha}} \mu^{\Dot{\beta}} \frac{\partial^2 f }{\partial \mu^{\Dot{\alpha}} \partial \mu^{\Dot{\beta}}})\\
   &\,= -2 (\widetilde{m}_\phi^2 -1) \phi \ .
\end{aligned}
\end{equation}
%%%%%%%%%%%%%%%%%%%%
\section{Spacetime and twistor space as homogeneous spaces} \label{app:homogeneous}
Complex hyperbolic space and its minitwistor space can be regarded as homogeneous spaces; more specifically, quotient spaces of the Lie group $G\equiv\sltc\times\sltc$, which is the isometry group of complex hyperbolic space, by a closed subgroup (the stabiliser). This serves to establish a more rigorous treatment of the structures and results we have given above, such as the space of homogeneous functions on twistor space and the Penrose transform. We shall closely follow the treatment given in \cite{tsai}. In particular we have
\begin{equation}
    \h = \frac{\sltc\times\sltc}{\sltc\times \mathbb{Z}_2}\,,\quad \MT = \frac{\sltc\times\sltc}{P\times P^{t}}\,,
\end{equation}
where the quotienting group in the latter case, $H\equiv P\times P^{t}\subset \sltc\times\sltc$ is given by,
\begin{equation}
    P \times P^{t}= \left\{\left(\begin{pmatrix}
        a&&b\\
        0&&a^{-1}
    \end{pmatrix},\begin{pmatrix}
        {a'}^{-1} & 0 \\
        -{b'} & a'
    \end{pmatrix}\right),\, a,a'\in\C^{*},\, b,b'\in\C
    \right\}\,.
\end{equation}

Hyperbolic space can be expressed in this way because the spinor equivalent of the coordinate $x^\ada$ is itself an element of $\sltc$ by virtue of the condition $\det(x^\ada)=x\cdot x=1$. The quotienting subgroup $\sltc\times \mathbb{Z}_2$ may be embedded into $G$ by $(g,\pm)\mapsto(g,\pm {g^t}^{-1})$ and so the quotient space is given by the equivalence classes of $G$ under
\begin{equation}
    (g_1,g_2)\sim(g_1 g, \pm g_2 {g^{t}}^{-1})\,,
\end{equation}
with $(g_1,g_2)\in G$ and $g\in\sltc$. Then, one can always choose $g=g_2^{t}$ and the positive sign in the second component to obtain the representative $(g_1 g_2^{t},\mathds{1})$ which can be identified to a point in hyperbolic space with spinor coordinate $x=g_1 g_2^{t}$. Note, that the choice of $\pm$ sign in the second component is precisely the $x\xr-x$ equivalence that was accounted for earlier in Eq. \eqref{hyperbolic}.

Minitwistor space on the other hand involves quotienting by the subgroup $H$, whose action on $(g_1,g_2)\in G$ is
\begin{equation}
    (g_1,g_2)=\left(\begin{pmatrix}
        \alpha & \beta\\
        \gamma & \delta
    \end{pmatrix},\begin{pmatrix}
        \epsilon & \varphi\\
        \mu & \nu
    \end{pmatrix}\right)\mapsto  \left(\begin{pmatrix}
        a\alpha &\, a^{-1}\beta+b\alpha\\
        a\gamma &\, a^{-1}\delta+b\gamma
    \end{pmatrix},\begin{pmatrix}
        {a'}^{-1}\epsilon-b'\varphi &\, {a'}\varphi\\
        {a'}^{-1}\mu-b'\nu &\, {a'}\nu
    \end{pmatrix}\right)\,.
\end{equation}
We can identify the quotient space as before, by choosing representatives for the cosets $(g_1,g_2)H$. Taking the case where $\alpha\neq0$ and $\epsilon\neq0$ we can choose $a=\alpha^{-1}$, $b=-\beta$, $a'=\varphi^{-1}$ and $b'=\epsilon$ to obtain
\begin{equation}
    [(g_1,g_2)H] = \left(\begin{pmatrix}
        1 &0\\
        \gamma/\alpha &\, 1
    \end{pmatrix},\begin{pmatrix}
        0 &\,1\\
        -1 &\, \nu/\varphi
    \end{pmatrix}\right)
\end{equation}
As the two matrices $g_{1,2}$ are of unit determinant we have $v_{1}\equiv(\alpha,\gamma)\neq(0,0)$ and $v_2\equiv(\varphi, \nu)\neq(0,0)$, and so $v_{1,2}$ can be considered to be projective coordinates on two copies of the Riemann sphere. This interpretation is justified as scaling either vector by a non-zero complex number, $v_{i}\xr\lambda_i v_i$ does not affect the ratios $\gamma/\alpha$ and $\nu/\varphi$ and so describes the same equivalence class. A similar argument follows for the case when $\alpha=0$ and/or $\epsilon=0$, in which case different choices of $a,b,a',b'$ must be taken. This is equivalent to using a different coordinate chart on the Riemann spheres which do not include the coordinate singularities. Hence
\begin{equation}
    \MT = \frac{\sltc\times\sltc}{P\times P^{t}}\cong \cp^1\times \cp^1\,.
\end{equation}
Physically, one can think of $H\equiv P\times P^{t}$ as the isometries of the totally geodesic null
hypersurfaces in $\h$.

Understanding the descriptions of spacetime and twistor space as homogeneous spaces also sheds light on the homogeneous `twistor functions' that are used as cohomology representatives in the Penrose transform. These complex-valued twistor functions, when defined at a given point form a vector space $V$ under addition, thus defining them over minitwistor space gives rise to an homogeneous vector bundle. Starting with a vector bundle in $G$, the homogeneous vector bundle can be obtained by enforcing the equivalence relation
\begin{equation}\label{fibereq}
    (g,v)\sim(gh,\rho(h^{-1})v)
\end{equation}
where $g\in G$, $v\in V$, and $\rho$ is a complex one-dimensional representation of $H$ given by
\begin{equation}
    \rho\left(\left(\begin{pmatrix}
        a&&b\\
        0&&a^{-1}
    \end{pmatrix},\begin{pmatrix}
        {a'}^{-1} & 0 \\
        -{b'} & a'
    \end{pmatrix}\right)\right) = a^{-m}{a'}^{-n}\,.
\end{equation}
This leads to the quotient space $(G\times V) /\sim$ with a natural projection map $\hat\pi:(G\times V)/\sim\,\,\xr G/H\cong\MT$, which takes equivalence classes $[(g,v)]$ to the corresponding left cosets
\begin{equation}
    \hat\pi: [(g,v)]\mapsto gH\,.
\end{equation}
The space $(G\times V)/\sim$ along with the map $\hat\pi$ can be shown to constitute a homogeneous vector bundle associated to the representation $\rho$, also known as the holomorphic line bundle $\cO(n\,|\,m)$ over $\MT$ \cite{Ward:1990vs}. This explains the scalings defined in Eq.~\eqref{scaling}.

\section{An overview of Čech versus Dolbeault representatives} \label{ap:rep}

The Penrose transform states that solutions to the equations of motion are given by \textit{sheaf cohomology group} representatives on twistor space. In the main text we have used representatives of Čech cohomology groups which are a suitable approximation to those of sheaf cohomology, however rather often in the literature an equivalent cohomology known as \textit{Dolbeault} cohomology is used. In this Appendix we briefly summarise some of the practical differences and connections between \cech and Dolbeault representatives.\\

Once the incidence relations are applied, the integral in Eq. \eqref{penrosetransform} is performed over a closed contour on the Riemann sphere. This contour separates the sphere into two regions, each containing singularities (else the integral would vanish). By Cauchy's theorem, this implies that Eq. \eqref{penrosetransform} is invariant under (now denoting $f$ by $f^C$ to denote Čech)
\begin{equation}\label{eq:cechequiv}
    f^C \xrightarrow{}{f'}^{C}= f^C +h_0 - h_1,
\end{equation}
where $h_0$ is holomorphic on one side of the contour and $h_1$ is holomorphic on the other. The motivation behind identifying the twistor functions $f^C$ in the Penrose transform with \cech cohomology representatives is well summarised in \cite{Chacon:2021lox} and references therein. Essentially, each term in Eq. \eqref{eq:cechequiv} can be understood as $p$-cochains, which are functions defined on the intersections of open sets $\{U_i\}$ of the Riemann sphere parametrised by $\lambda_\alpha$. These $p$-cochains have $p$ antisymmetric indices which indicate which intersection they are defined upon (e.g. $f_{01}$ is defined on the intersection of open sets $U_0\cap U_1$) and the degree of a $p$-cochain can be raised by a \textit{coboundary} operator $\delta_p$ to obtain a $(p+1)$-\textit{coboundary}. If $\delta_p$ annihilates a $p$-cochain, then the cochain is designated as a \textit{cocycle}. The \cech cohomology group is the group (under addition) of cocycles, modulo the group of coboundaries, that is, the cocycles that differ by the addition of a coboundary are equivalent \cite{Nakahara:206619}.

Hence the redundancy in $f^C$ precisely reflects the statement that $f^C$ and ${f'}^{C}$ are both representatives of the same \cech cohomology class, and by the Penrose transform correspond to the same spacetime solution.

On the other hand, the Dolbeault cohomology starts by noticing that on a complex manifold with local complex coordinates $(z^A,\bar{z}^{\bar A})$, $k$-forms can be decomposed into a sum of so called $(p,q)$-forms which carry $p$ down-indices corresponding to the holomorphic coordinates $z^A$ and $q$ down-indices corresponding to the anti-holomorphic $\bar{z}^{\bar A}$ coordinates. It follows that the exterior derivative splits as
\begin{equation}
    \text{d} = \partial + \Bar{\partial},
\end{equation}
where  $\partial^2 = \Bar{\partial}^2 =0$ (nilpotency). The operator $\bar{\del}$, also known as the Dolbeault operator, annihilates holomorphic functions on the manifold, i.e.  $\bar{\del}f=0$ if $f$ is holomorphic. Due to the nilpotency of $\dbar$, we may add a $\dbar$-exact function, i.e. one of the form $h=\dbar g$ to any holomorphic (or $\dbar$-closed) function to obtain another $\dbar$-closed/holomorphic function. The Dolbeault cohomology groups $H^{p,q}$ can then be defined as the spaces of $(p,q)$-forms $f^{D}$ that are $\dbar$-closed but not $\dbar$-exact, modulo the equivalence relation
\begin{equation}\label{eq:dolbeault}
    f^{D}\sim f^{D}+\dbar g\,,\quad f^{D}\in\Omega^{p,q},\,g\in\Omega^{p,q-1}\,,
\end{equation}
where $\Omega^{p,q}$ is the space of $(p,q)$-forms.

For the purposes of the Penrose transform the relevant Dolbeault cohomology group is $H^{0,1}$ and the Penrose transform takes the slightly different form
\begin{equation}\label{eq:formpenrose}
    \Psi_{\alpha_1\ldots\alpha_{2s}}(X)=\frac{1}{2\pi\ri}\int_X\la\lambda\rd\lambda\ra\,\wedge \lambda_{\alpha_{1}}\ldots\lambda_{\alpha_{2s}}\,f^{D}(\mu_X,\lambda)\,,
\end{equation}
where now the integral is over the entire Riemann sphere $X$ as opposed to a closed contour. In accordance with this, the representative $f^D$ and integral measure are $(0,1)$ and $(1,0)$-forms respectively, such that their exterior product gives a top form that can be integrated over $X$. By usage of Stokes' theorem, it can be shown that the addition of a $\dbar$-exact form to $f^D$ leaves the integral \eqref{eq:formpenrose} unchanged and so does not affect the spacetime solution. This is precisely the equivalence in \eqref{eq:dolbeault} and so justifies the usage of Dolbeault cohomology classes to represent distinct spacetime solutions. Finally, note that the Dolbeault representatives must also respect the same homogeneity/scaling properties as their \cech counterparts, with respect to the twistor space coordinates. For more details see \cite{Adamo:2017qyl}.\\

If Dolbeault and \cech cohomologies are indeed equivalent in our context, there should be a clear way to connect representatives in either language to each other. As reviewed in \cite{Chacon:2021lox} and references therein, one can start by defining an atlas on $X$ with two charts $U_{0,1}$ such that $X=U_0\cup U_1$ and $f^C$ is holomorphic on the intersection $U_0\cap U_1$. Let $\gamma$ (the integration contour in the \cech version of the Penrose transform) lie entirely inside $U_0\cap U_1$. Then the equivalent Dolbeault representative of $f^C$ in the patch $U_0$ is given by
\begin{equation}\label{eq:cechtodolb}
    f^D = \begin{cases}
    \dbar f^C &\quad\text{in}\quad U_0\,,\\
    0 &\quad\text{in}\quad U_1 \,.
    \end{cases}
\end{equation}
To see that this is the correct Dolbeault representative we can insert the above into the Penrose transform \ref{eq:formpenrose} to find
\begin{equation}
\begin{aligned}
        \Psi_{\alpha_1\ldots\alpha_{2s}}(X)&=\frac{1}{2\pi\ri}\int_{U_0}\la\lambda\rd\lambda\ra\,\wedge \lambda_{\alpha_{1}}\ldots\lambda_{\alpha_{2s}}\,\dbar f^C|_X(\mu_X,\lambda)\\
        &=\frac{1}{2\pi\ri}\oint_{\gamma}\la\lambda\rd\lambda\ra \lambda_{\alpha_{1}}\ldots\lambda_{\alpha_{2s}}f^C(\mu_X,\lambda)\,,   
\end{aligned}
\end{equation}
where in the second line we have used Stokes' theorem and the fact that the boundary contour of $U_0$ can be deformed to $\gamma$ as $f^C$ is holomorphic on $U_0\cap U_1$. Note that the Dolbeault representative cannot be globally defined as $\dbar f^C$ as this would be $\dbar$ exact and so give a trivial Penrose transform. \\

\textbf{Example -- negative helicity plane waves:} For instance, as shown in \cite{Chacon:2021lox}\footnote{Note that to go from \cite{Chacon:2021lox} to the current conventions, one should take $\pi_{A'}\xrightarrow{} \lambda_{\alpha}$, $\omega_{A} \xrightarrow{} \mu_{\dot{\alpha}}$.}, in four spacetime dimensions helicity $-s <0$ plane wave solutions of the form
\begin{equation}\label{eq:plane}
    \phi_{\alpha_{1}...\alpha_{2s}}(X)= p_{\alpha_{1}}...p_{\alpha_{2s}} e^{\ri p \cdot X}\,,
\end{equation}
with $p_{\alpha}\tilde{p}_{\dot{\alpha}}$ being the spinor equivalent of the momentum four-vector, have the Dolbeault representatives
\begin{equation}
    f^{D}_h=\left(\frac{\braket{ap}}{\braket{a \lambda}}\right)^{1+2s}\bar{\delta}(\braket{p\lambda })\,\exp\left[\ri\frac{\braket{ap}}{\braket{a\lambda}}[\mu \tilde{p}]\right],
\end{equation}
where $a_{\dot{\alpha}}$ is an arbitrary constant spinor. The object $\bar{\delta}(u)$ is a $(0,1)$-form which behaves like a Dirac delta function when integrated against holomorphic $(1,0)$-forms; it is defined as $\bar{\delta}(u)\equiv\dbar(1/u)$. Using the Schouten identity as well as projective scaling of $\lambda_\alpha$, we can write $\lambda_\alpha = b_\alpha + z a_\alpha$ where $b_\alpha$ is some constant spinor satisfying $\la ab\ra\neq0$. Then the above reduces to
\begin{equation}
    f^{D}_h=\left(\frac{\braket{ap}}{\braket{a b}}\right)^{1+2s}\frac{1}{\braket{pa}}\,\dbar\left(\frac{1}{\frac{\braket{bp}}{\braket{ap}}+z}\right)\,\exp\left[\ri\frac{\braket{ap}}{\la ab\ra}[\mu \tilde{p}]\right]\,.
\end{equation}
When the incidence relations are applied, the coordinate $z$ parametrises a patch on the Riemann sphere $X$ (note that we use the same label $X$ for the spacetime coordinate and the Riemann sphere, as the incidence relation maps points in spacetime to embedded Riemann spheres in twistor space) that excludes the point $\lambda_\alpha\propto a_\alpha$. The $z$ dependence outside of the $\dbar$ disappears and so the \cech representative is given by
\begin{equation}\label{eq:cechwave}
    f^{C}_h=\left(\frac{\braket{ap}}{\braket{a b}}\right)^{1+2s}\frac{1}{\braket{pa}}\left(\frac{1}{\frac{\braket{bp}}{\braket{ap}}+z}\right)\exp\left[\ri\frac{\braket{ap}}{\la ab\ra}[\mu \tilde{p}]\right]\,.  
\end{equation}

It is straightforward to verify that inserting this into the \cech Penrose transform (including using the incidence relations) with the $\gamma$ contour enclosing the pole at $z=\la pb\ra/\la ap\ra$ gives Eq. \eqref{eq:plane}. The representative \eqref{eq:cechwave} is given as a function of the $z$ coordinate, but we may revert it to an expression in terms of the twistor variable $\lambda_\alpha$ to obtain
\begin{equation}
    f^C_h(\mu^{\dot\alpha},\lambda_\alpha) = \left(\frac{\braket{ap}}{\braket{a \lambda}}\right)^{1+2s}\frac{1}{\braket{p\lambda}}\exp\left[\ri\frac{\braket{ap}}{\la a\lambda\ra}[\mu \tilde{p}]\right]\,,
\end{equation}
making its scaling weight of $-2-2s$ manifest.

%%%%%%%%%%%%%%%%%%%%%
\section{Dimensional reduction via the Mellin transform} \label{app:dimred}
Cohomology representatives for minitwistor space (which have independent scaling weights for $\mu^{\dot\alpha}$ and $\lambda_\alpha$) can be obtained from cohomology representatives of `4d' twistor space (corresponding to four-dimensional flat spacetime) via the Mellin transform \cite{Pasterski:2017kqt,Sharma:2021gcz}
\begin{equation}\label{reduction}
    f^{3d}(\mu,\lambda)=\int_0^{\infty}\frac{\rd\omega}{\omega}\omega^{\Delta-s}f^{4d}(\omega\mu,\lambda)\,.
\end{equation}
In the above equation either \cech or Dolbeault representatives may be used, and the scaling weight of $f^{4d}$ is $-2-2s$.\\

\textbf{Example -- plane waves: } In \cite{Bu:2023cef} it is shown that the Mellin transform of a scalar momentum eigenstate (plane wave) representative gives a minitwistor representative that maps to a scalar bulk-to-boundary propagator on spacetime under the Penrose transform. It is straightforward to generalise this result to plane wave solutions of non-zero negative helicity. Taking the \cech representative for the negative helicity plane wave
\begin{equation}
    f^{4d}=\left(\frac{\braket{ap}}{\braket{a b}}\right)^{1+2s}\frac{1}{\braket{pa}}\left(\frac{1}{\frac{\braket{bp}}{\braket{ap}}+z}\right)\exp\left[\ri\frac{\braket{ap}}{\la ab\ra}[\mu \tilde{p}]\right]\, ,  
\end{equation}
and plugging into the Mellin transform above gives an integral that converges provided that $\text{Im}\left(\braket{ap}\braket{a\lambda}^{-1}[\mu\tilde p]\right)>0$ and $\text{Re}(\Delta+s)>0$, to give
\begin{equation}
    f^{3d}(\mu,\lambda) =-i^{\Delta+s}\Gamma(\Delta+s)\braket{ap}^{s-\Delta}\braket{ab}^{\Delta-1-s}[\mu \tilde p]^{-\Delta-s}\left(\frac{1}{\frac{\braket{bp}}{\braket{ap}}+z}\right)\,.
\end{equation}
In the above expression the $\lambda$ dependence is implicit in the $z$ and $b$ dependence, but can be restored by eliminating $b_\alpha=\lambda_\alpha-z a_\alpha$ to find
\begin{equation}
    f^{3d}(\mu,\lambda) =i^{\Delta+s}\Gamma(\Delta+s)\braket{ap}^{s-\Delta+1}\braket{a\lambda}^{\Delta-1-s}[\mu \tilde p]^{-\Delta-s}\left(\frac{1}{\braket{p\lambda }}\right)\,,
\end{equation}
which shows the homogeneity of the 3d representative is $\cO(-\Delta-s,\Delta-s-2)$. The Penrose transform then implies that this representative corresponds to a solution of the three-dimensional spacetime equations of motion \eqref{nonzerospin} with $\tilde m = \Delta-1$.
Performing the Penrose transform gives
\begin{equation}
    \phi_{\alpha_1\ldots\alpha_{2s}}(X)=i^{\Delta+s}\Gamma(\Delta+s) p_{\alpha_1}\ldots p_{\alpha_{2s}}\frac{1}{(X\cdot p)^{\Delta+s}}\,.
\end{equation}
In the case of plane wave solutions, the Mellin transform can be used to produce 3d minitwistor representatives of different $\tilde m$ (or equivalently, different conformal dimension $\Delta$).\\

In the previous plane wave example, the Mellin transform allowed one to obtain a family of representatives parametrised by the value of $\Delta$ which is related to the $\tilde m$ parameter of the 3d solution. In particular, massive 3d representatives were obtained from the massless 4d ones. This however is not possible in general. For instance, for representatives of the form (a notable example being Taub-NUT black holes \cite{Chacon:2021wbr})
\begin{equation}
    f^{4d}(\mu,\lambda)= \left(Q^{\alpha}_{\Dot{\beta}}\lambda_{\alpha} \mu^{\Dot{\beta}}\right)^{-m}
\end{equation}
the integration parameter $\omega$ in the Mellin transform factorises out of the twistor function and the three dimensional representative is a constant multiplying the four dimensional one. Hence, while a 3d representative is obtained, this corresponds to the massless case. In other words, the dimensional reduction does not give rise to the representatives of warped AdS constructed in Section \ref{sectiontyped}.\\

Another interesting relationship between 3d and 4d representatives can be obtained as follows. Considering the product of 3d representatives with oppositve helicity gives 
\begin{equation}
    f_{\tilde{m}_{+,s}}(\alpha' \mu,\alpha \lambda)f'_{\tilde{m}_{-,s}}(\alpha' \mu,\alpha \lambda)=\alpha^{-2-2s}\alpha^{\prime -2-2s}f_{\tilde{m}_{+,s}}(\mu,\lambda)f'_{\tilde{m}_{-,s}}(\mu,\lambda),
\end{equation}
where $\tilde m_{\pm,s}= \pm(\tilde{m}_0+s)$. In other words, we see that $\lambda$ and $\mu$ have now identical scalings and that the mass dependence drops. The homogeneity on the RHS corresponds to the scaling of two spin $s$, massless, 4d twistor representatives and we have
\begin{equation}
    f^{3d}_{\tilde{m}_{+,s}}(\mu,\lambda)f^{3d \prime}_{\tilde{m}_{-,s}}(\mu,\lambda)= (f^{4d}_s(\mu,\lambda))^2.
\end{equation}
For instance, one can take the simple following example
\begin{equation}
\begin{aligned}
     f^{3d}_{\tilde{m}_{+,s}} = \braket{\lambda A}^{\tilde m_0-1}[\mu B]^{-\tilde{m}_0-2s -1}, \\
     f^{3d \prime}_{\tilde{m}_{-,s}} = \braket{\lambda A}^{-\tilde{m}_0-2s -1}[\mu B]^{\tilde m_0-1}, \\
     f^{4d}_s(\mu,\lambda)= (\braket{\lambda A}[\mu B])^{-s-1}.
\end{aligned}     
\end{equation}

\section{Petrov classification}\label{app:petrov}
In 3 dimensions, the Cotton spinor $C_{\alpha\beta\gamma\delta}$ and the traceless Ricci spinor $S_{\alpha\beta\gamma\delta}$ can be used for algebraic classifications of spacetimes since both are traceless and symmetric \cite{Milson:2012ry,castillo20033,doi:10.1063/1.1592611}. In TMG, The Cotton spinor is proportional to the traceless Ricci so that the classification is equivalent, but this is not the case in general. Given that the Cotton spinor has 4 indices, all the symmetric possibilities can be listed up to overall spacetime dependent factor as follows
\begin{equation}\label{petrov}
    \begin{aligned}
       \text{type I  }\phantom{aaa} C_{\alpha\beta\gamma\delta} &\,=\psi(x) r_{(\alpha} s_{\beta} t_{\gamma} u_{\delta)}, \\
        \text{type II   }\phantom{aaa} C_{\alpha\beta\gamma\delta} &\,=\psi(x) r_{(\alpha} r_{\beta} s_{\gamma} t_{\delta)}, \\
       \text{type D   }\phantom{aaa} C_{\alpha\beta\gamma\delta} &\,=\psi(x) r_{(\alpha} r_{\beta} t_{\gamma} t_{\delta)}, \\
       \text{type III   }\phantom{aaa} C_{\alpha\beta\gamma\delta} &\,=\psi(x) r_{(\alpha} r_{\beta} r_{\gamma} s_{\delta)}, \\
      \text{type N   }\phantom{aaa}  C_{\alpha\beta\gamma\delta} &\,=\psi(x) r_{(\alpha} r_{\beta} r_{\gamma} r_{\delta)}, \\
     \text{type O   } \phantom{aaa}  C_{\alpha\beta\gamma\delta} &\,=0.
    \end{aligned}
\end{equation}
where $r_{\alpha}, s_{\alpha}, t_{\alpha}, u_{\alpha}$ are non-proportional spinors called principal spinors. Note that some types include other types: we have the chains type O $\subset$ type N $\subset$ type D $\subset$ type II $\subset$ type I and type O $\subset$ type N $\subset$ type III $\subset$ type II $\subset$ type I. 

The interpretation of these spacetimes is as follows. type O spacetimes are conformally flat, type N are associated to a transverse gravitational wave (such as PP waves), type D describe isolated massive objects (such as black holes), type III are associated to longitudinal gravitational waves and type I are the most generic types of spacetime. We will consider solutions around fixed backgrounds, hence we will pay special attention to Kerr-Schild spacetimes (a special case of type II) as the statements we make about them will be exact.\footnote{This is because, writing $g=\bar{g}_{\mathbb{H}_3}+h$, the Cotton tensor is linear in $h$ so that $\nabla_gC=\nabla_{\mathbb{H}_3}C$ is true exactly whereas it will only be true at linear order for type I} 

Since $r^{\alpha}r_{\alpha}=0$, it is easy to find twistor representatives corresponding to each type. Indeed, as explained in \cite{Penrose:1986ca}, a twistor representative with a single pole of order $2s+1 -J$ enclosed inside the contour corresponds to a field with a $J$-fold principal spinor. In particular, type N and type II fields representatives will respectively have one simple pole and one third order pole enclosed inside the contour. Type D fields have a pair 2-fold principal spinors, which require two poles of order 3 on either side of the contour, which forces $\widetilde m$ to take only three possible values. One could naively expect that fine-tuned coefficient might exist such that when contracted, the different poles cancel each other. However, as we show in Section \ref{sectiontyped}, this does not happen and only the three values $\widetilde m=\pm 3,0$ are allowed.

\newpage

%%%%%%%%%%%%%%%%%%%%
\bibliography{refs.bib}
\bibliographystyle{JHEP}

\end{document}